# REFORMULATION INSTEAD OF RENORMALIZATIONS

## Vladimir Kalitvianski

vladimir.kalitvianski@wanadoo.fr
CEA/Grenoble, 2008

In this article I show why the fundamental constants obtain perturbative corrections in higher orders, why the renormalizations "work" and how to reformulate the theory in order to avoid these technical and conceptual complications. I demonstrate that the perturbative mass and charge corrections are caused, roughly speaking, with the *kinetic* nature of the interaction Lagrangian. As it is not purely quantum mechanical (or QFT) specific feature, the problem can be demonstrated on a classical two-body problem. The latter can be solved in different ways, one of them being correct and good for applying the perturbation theory (if necessary) and another one being tricky and awkward. The first one is physically and technically natural – it is a center-of-inertia-and-relative-variable formulation. The second one – mixed variable formulation – is unnecessarily complicated and leads to the mass and charge corrections even in the Newtonian mechanics of two bound bodies. The gauge theories are factually formulated in the mixed variables - that is why they bring corrections to the fundamental constants. This understanding opens a way of correctly formulating the gauge QFT equations and thus to simplify the calculations technically and conceptually. For example, in scattering problems in QED it means taking exactly into account the quantized electromagnetic field influence in the "in" and "out" states of charged particles so no infrared and ultraviolet problems arise. In bound states it means obtaining the energy corrections (the Lamb shift, the anomalous magnetic moment) quite straightforwardly and without renormalizations.

**INTRODUCTION**

The interaction term that causes the mathematical and conceptual problems is the so-called self-action term $L_{int} = j \cdot A_{tot}$. The problem of self-action of elementary particles is known from the Classical Electrodynamics (CED). H. Lorenz was the first who worried about the energy-momentum conservation in electrodynamics of elementary electron [1]. He thought that the only way of automatically taking into account the radiative losses in the electron dynamics would be putting the proper particle field (including the radiated one) into the right-hand side of the electron (Newton) equations (Lorentz or self-action ansatz). He expected a small term that would slightly decrease the electron acceleration due to "radiative friction". In fact there is another way of preserving the conservation laws, but we first consider this question in its historical aspect.

But before that, knowing that the radiation is caused with the charge acceleration $\ddot{\mathbf{r}}$ (not a function of $\mathbf{r}$ or $\dot{\mathbf{r}}$) and applying a purely *phenomenological* approach to describing the electron dynamics, we might modify the Newton equation in the following way:

$$m_e \ddot{\mathbf{r}} = \mathbf{F}_{ext} - \delta m \cdot \ddot{\mathbf{r}}. \qquad (\mathbf{I}1)$$

As far as the radiative losses are experimentally small, the dimensional coefficient $\delta m$ is expected to be small too. It would be quite relevant to call then $\delta m$ the "radiative added mass". In experiments with sufficiently strong accelerations the radiative added mass $\delta m$ is just added to $m_e$, so the resulting acceleration $\ddot{\mathbf{r}} = \mathbf{F}_{ext}/(m_e + \delta m)$ and velocity $\dot{\mathbf{r}}$ would be somewhat smaller due to radiative losses. On the other hand, in a *static* experiment in a gravitational field, when one "weights" the electron, one obviously deals with a pure "mechanical" $m_e$, not with a sum. In a "weighting" experiment no acceleration in the equations appears so the gravitational force, say $m_e g$ for simplicity, is counterbalanced with an elastic (or electrostatic $eE$) force, for example: $(m_e g - e\mathbf{E}) = 0$. Thus, $m_e$ and $\delta m$ can be measured and treated *separately*. Naturally, the kinetic energy and the momentum of a *free* electron are expressed via $m_e$ solely (no radiation reaction effect should be involved), so the term $\delta m$ is *not always* added to $m_e$ - they are *distinguishable*. It becomes especially evident in a moving reference frame. For example, the heat released in matter with a fast projectile is equal to its initial kinetic energy $m_e v^2/2$. This would be quite a good physical approach - without conceptual and mathematical difficulties, just in the spirit of the phenomenological description of the usual mechanical friction, if it worked, of course. Alas, it does not work. Adding a term $-\delta m \cdot \ddot{\mathbf{r}}$ does not help conserve the energy-momentum. It is rather clear in case of a uniform magnetic field or in case of a potential external force – the particle energy does not vary as the radiative losses $\propto \int_0^t dt' \ddot{\mathbf{r}}^2(t')$, but differently: $E(t) - E(0) = -\delta m \left[ \dot{\mathbf{r}}^2(t) - \dot{\mathbf{r}}^2(0) \right]$. Hence we need some other ideas to derive the radiation reaction force.

Unfortunately all attempts to derive it theoretically have failed. H. Lorentz, proceeding from the classical notion of an electron as a *self-acting* charge distribution with density $\rho(\mathbf{r})$, obtained the very same term $\delta m \cdot \ddot{\mathbf{r}}$ as in (**I**1) with $\delta m$ as a function of the charge size $r_e$: $\delta m \propto 1/r_e$. I must say that calculating the forces acting between different parts of a distribution $\rho(\mathbf{r})$ and supposing at the same time that these forces do not provide any internal dynamics is rather strange. As to the electron dynamics as a whole, one factually thus calculates a self-induction effect – resistance to any change of the state of a uniform motion $\dot{\mathbf{r}} = \text{const}$. It is not a radiation reaction force because it does not help conserve correctly the energy-momentum (as we have seen above) and it is highly size- (or model-) dependent. Hence, we conclude that although of a good dimension, the self-action term $L_{int} = j \cdot A_{tot}$ is of a bad functional dependence: $-\delta m \cdot \ddot{\mathbf{r}}$ is not what we wanted.

Following this calculation, can we pretend that a self-induction mechanism is at least responsible for the electron inertia? No, I am afraid, we cannot. The problem of stability of such a stiff charge distribution requires an explanation. H. Poincaré admitted a non-pointlike structure of the electron charge and counterbalanced the electrostatic repulsion forces with non-electromagnetic ones, leaving clarifying the nature of the latter to the future [2]. These non-electromagnetic forces may have their own "self-induction" and "radiation resistance" contributions. But Lorenz totally ignored them and made $r_e$ tend to zero. As a result, his value of $\delta m$, called often the electromagnetic mass, went to infinity (too strong self-induction). The nice Noether theorem about conservation of the total (particle + field) energy-momentum in the self-acting CED fails in practice for a point-like electron precisely because of this infinite term. What is a use of any formal conservation law if no dynamics can be in reality calculated ($\dot{\mathbf{r}} = \text{const}$)? (This theorem holds even for non-physical solutions.)

A finite "remainder" turned out then to be proportional to the third derivative of electron coordinate $\mathbf{r}$:

$$m_e \ddot{\mathbf{r}} = \mathbf{F}_{ext} - \delta m \cdot \ddot{\mathbf{r}} + \frac{2e^2}{3c^3} \dddot{\mathbf{r}}. \quad (\mathbf{I}2)$$

To overcome the failure of this calculation, an additional "physical idea" was advanced: to consider the original Newton equation as already containing the "self-induction" effect $\delta m \cdot \ddot{\mathbf{r}}$, but not containing anything from the jerk term $\frac{2e^2}{3c^3} \dddot{\mathbf{r}}$. According to it, we must keep in (**I**2) the term $\frac{2e^2}{3c^3} \dddot{\mathbf{r}}$ solely and discard $\delta m \cdot \ddot{\mathbf{r}}$. Often it is represented as adding $\delta m$ to a "bare" mass as if we had a Newton equation with a "bare" mass $m_B$ before our development (**I**2). As it is not the case, these "physical ideas" are not convincing to me. Moreover, I can safely say there are no electromagnetic and bare masses at all since these two "physical concepts" cancel each other and disappear from CED equations.

It is obvious that this reasoning results in fact in *discarding* forever the term $\delta m \cdot \ddot{\mathbf{r}}$ from the equation (**I**2) with a hope that the "remainder" (a jerk term) provides a good description of the radiation resistance:

$$m_e \ddot{\mathbf{r}} = \mathbf{F}_{ext} + \frac{2e^2}{3c^3} \dddot{\mathbf{r}}. \quad (\mathbf{I}3)$$

That means *postulating* this new (trial) equation (**I**3) as an equation for the electron dynamics. The term $\frac{2e^2}{3c^3} \dddot{\mathbf{r}}$ has another functional dependence.

The successive analysis showed, however, that the new equation had not become better physically and mathematically – its exact solution is not physical at all (a runaway solution). In addition, the third order equation needs a third initial constant.

It is possible to eliminate the runaway solutions by the price of acausal description - the present particle behavior is then determined with (generally unknown) future [3]. The measured initial acceleration, as a third initial constant, is generally different from the conditions on $\ddot{\mathbf{r}}(0)$ determined with future forces. Fortunately, such an approach is not (at least, widely) accepted by the physical community. Hence, we conclude that although of a good dimension, the jerk term $\propto \dddot{\mathbf{r}}$ is of a bad functional dependence, i.e., it is not what we wanted. So, further modifications of (**I3**) are necessary.

It is now widely accepted to replace the jerk term with the force derivative [4]:

$$m_e \ddot{\mathbf{r}} = \mathbf{F}_{ext} + \frac{2e^2}{3m_e c^3} \dot{\mathbf{F}}_{ext}(\mathbf{r}, \dot{\mathbf{r}}, t). \qquad (\mathbf{I4})$$

Note, the force derivative term in (**I4**) depends on unknown particle coordinates and velocities rather than on the zeroth-order solutions of Eq. (**I4**): $\dot{\mathbf{F}}_{ext} = \dot{\mathbf{F}}_{ext}(\mathbf{r},\dot{\mathbf{r}},t) \neq \mathbf{f}(t) = \dot{\mathbf{F}}_{ext}\left(\mathbf{r}^{(0)}(t), \dot{\mathbf{r}}^{(0)}(t), t\right)$. For example, in the oscillator equation $\dot{\mathbf{F}}_{ext} = \dot{\mathbf{F}}_{ext}(\mathbf{r},\dot{\mathbf{r}},t) \propto \dot{\mathbf{r}}$ is a damping force whereas $\dot{\mathbf{F}}_{ext}\left(\mathbf{r}^{(0)}(t), \dot{\mathbf{r}}^{(0)}(t), t\right) = \dot{\mathbf{r}}^{(0)}(t)$ is a resonant external force – first it damps the existing oscillations, and then it pumps them without limit. It means that Eq. (**I4**) is not a perturbative version of Eq. (**I3**), but is *another* trial equation requiring no third initial constant. It does not follow from (**I2**); it was luckily found to be physically (more or less) acceptable by trying different "terms of a good dimension". Generally, it does not provide the energy-momentum conservation, so it is still an approximate equation. F. Rohrlich considered it the best possible description in the frame of CED [5]. The term $\frac{2e^2}{3m_e c^3} \dot{\mathbf{F}}_{ext}(\mathbf{r},\dot{\mathbf{r}},t)$ is a third functional dependence tried for description the radiation reaction force. It means the mass renormalization in CED is not sufficient to obtain physically meaningful results. In other words, a self-acting CED is a non-renormalizable theory.

As we can see, nothing has left from the naive idea of taking into account the radiative losses in the electron dynamics with the self-action term $j \cdot A_{tot}$. We have passed through a series of failures with their unsatisfactory physical and mathematical "remedies". Explicitly covariant formulation of the electron dynamics [6] improved nothing in this respect. We must recognize that the idea of self-action has never been a success in CED. Despite that, this idea was adopted in the early QED (Dirac, Pauli, Heisenberg), apparently because of lack of other fruitful ideas in CED. No wonder the standard QED is also plagued with similar mathematical and conceptual problems. As far as the electron-field "coupling" is always treated in QED perturbatively, the exact solutions are unknown, so renormalizations are present in each perturbative order. It makes a false impression that the bare constants and perturbative corrections like self-induction and vacuum polarization are physically meaningful.

Besides, all the judgments are made basing on the perturbative solutions. Often the classical notions and solutions serve there as the physical "language" and the initial approximations. For example, it is seriously stated in QED that the infinite charge of the pointlike "bare" electron is screened due to infinite "vacuum polarization". In reality the classical "pointlike" charge is always the inclusive (= approximate, secondary, illusive) rather than an exact picture [7].

Is there another way of preserving the energy-momentum conservation laws (i.e., without self action)? Yes, there is one. Let us mention here that the free quantized electromagnetic field in QED is essentially considered as a set of elementary quantum oscillators which describe the field Fourier amplitudes. Then, if we look at the mechanical equations as at the center of inertia (CI) equations of a compound system and at the elementary oscillation equations as at the equations of relative (internal) motion of the compound system, then there is no need to put a "friction" term in the CI dynamics. This is the case for any macroscopic body, this is the case for atoms [7], and there are no physical reasons to consider the real electron differently. For that we must admit that the oscillator equations describe the relative motion in a compound system (called hereafter an electronium). Then the electric charge in the electronium is smeared quantum mechanically, described with form factors, and is rather different from the "pointlike" charge distribution. This simple and natural physical concept removes the mathematical problems in QED and therefore in CED. We are going to outline the arguments in the next chapters.

The question arises: why does the QED with the self-action and with renormalizations work then? The brief answer is – because of good luck. The self-action, as we could see and will see later, is a physically wrong idea, and the renormalization (discarding certain terms) is mathematically wrong too, but, roughly speaking, it removes the numerical wrongness introduced with the self-action effects (it is the only purpose of the renormalizations). But although the agreement with the experimental data is good, the renormalization philosophy ("vacuum polarization physics") is completely wrong; it misleads the researchers and does not give the true understanding of the physical reality. So the theory with renormalizations does not answer correctly to our physical questions. To see all this in details and to understand how to reformulate correctly the theory of interacting particles, let us consider first classical mechanical equations. Fortunately they contain *all* necessary features to demonstrate the self-action problem and its elimination. Transition to the quantum mechanical case is then elementary.

## 1. CLASSICAL THEORY OF TWO BOUND PARTICLES AND ITS PHYSICS

For that it will be necessary to present first some simple and very well known equations and discuss some obvious physics. Then the same problem will be formulated in the so-called mixed variables that resemble the QED variables.

Let us consider two classical bound particles ($\mathbf{r}_1, M_1$ and $\mathbf{r}_2, M_2$) in an external potential field $V_{ext.}$. This system may well model a complex material body since it has internal degrees of freedom. The external field is supposed to act only on the first particle: $\mathbf{F}_{ext} = \mathbf{F}_{ext}(\mathbf{r}_1)$. Until Chapter 3 we will consider a made-of-kit (mountable-dismountable) compound system with well-defined and known constants $M_1$, $M_2$, and $k$. The external force acts within a finite time interval $0 \leq t \leq t_F$. The inter-particle (or internal for short) force $\mathbf{F}_r(\mathbf{r}_1 - \mathbf{r}_2)$ will be assumed, when necessary, to be an oscillator-like for certainty. In fact our consideration is sufficiently general and is not limited to "bound" particles.

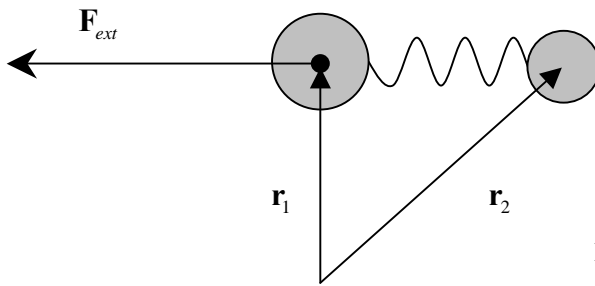

Fig. 1. Mountable-dismountable mechanical system.

The corresponding Newton equations are essentially coupled due to the force $\mathbf{F}_r$ arguments:

$$M_1 \dot{\mathbf{v}}_1 = \mathbf{F}_{ext}(\mathbf{r}_1) + \mathbf{F}_r(\mathbf{r}_1 - \mathbf{r}_2), \tag{1}$$

$$M_2 \dot{\mathbf{v}}_2 = -\mathbf{F}_r(\mathbf{r}_1 - \mathbf{r}_2). \tag{2}$$

From the first equation we see that the external field makes work on "accelerating" the first particle and on overcoming its "attraction" to the second one. Depending on the oscillation phase, the external force obtains "assistance" or "resistance" from the second particle.

The second equation says that the second particle is "attracted" by the first one. The words "attraction", "acceleration", etc., are used hereafter in an algebraic sense: they can be of any signs, of course.

Such a coupled equation system is not convenient to resolve the problem analytically, and some change of variables is in order.

### 1.1. Center of inertia and relative coordinates

The particle masses $M_1$ and $M_2$ may be called, to a certain extent, the "bare" masses, as if they belonged to independent (non-interacting) particles (at least it is so before assembling the system).

Let us introduce now the CI and relative coordinates (CIRC formulation):

$$\mathbf{r}_r = \mathbf{r}_1 - \mathbf{r}_2, \quad \mathbf{R}_{CI} = \frac{M_1 \mathbf{r}_1 + M_2 \mathbf{r}_2}{M_1 + M_2} = \mathbf{r}_1 - \frac{M_2}{M_{tot}} \mathbf{r}_r, \quad M_{tot} = M_1 + M_2, \quad \mu = \frac{M_1 M_2}{M_{tot}}. \tag{3}$$

They correspond now to two *quasi* particles (or subsystems) describing the centre of inertia and the relative motions (kind of collective variables). Due to linearity of the variable change, the quasi particle equations are also of the *second* order:

$$M_{tot} \ddot{\mathbf{R}}_{CI} = \mathbf{F}_{ext}\left(\mathbf{R}_{CI} + \frac{M_2}{M_{tot}} \cdot \mathbf{r}_r\right), \tag{4}$$

$$\mu \ddot{\mathbf{r}}_r = \mathbf{F}_r(\mathbf{r}_r) + \frac{M_2}{M_{tot}} \cdot \mathbf{F}_{ext}\left(\mathbf{R}_{CI} + \frac{M_2}{M_{tot}} \cdot \mathbf{r}_r\right). \tag{5}$$

The quasi particle masses $M_{tot}$ and $\mu$ may be called the "dressed" masses as if they belonged to some particles resulted from interaction of "bare" particles.

We define now the following natural dimensionless "coupling" constants:

$$\boxed{\varepsilon = \frac{M_2}{M_{tot}}}, \quad 0 < \varepsilon < 1, \quad \boxed{\varepsilon' = \frac{M_2}{M_1}}, \quad 0 < \varepsilon' < \infty, \tag{6}$$

$$\frac{1}{M_{tot}} = \frac{1}{M_1} \frac{1}{(1+\varepsilon')}, \quad \mu = M_2 \frac{1}{(1+\varepsilon')}, \quad \varepsilon = \varepsilon' \frac{1}{(1+\varepsilon')}. \tag{7}$$

I call $\varepsilon$ a "coupling" constant because it "couples" two equations. It is much similar to the electron charge $e$, which also couples the mechanical and the wave equations. When $\varepsilon = 0$ the equations become independent, decoupled. The "coupling" parameter $\varepsilon$ naturally appears in the exact solutions in the CIRC formulation along with $M_{tot}$ and $\mu$ as well as in a very important expression:

$$\mathbf{r}_1 = \mathbf{R}_{CI} + \varepsilon \cdot \mathbf{r}_r. \tag{8}$$

The corresponding Lagrangian is:

$$L_{CIRC} = \frac{M_{tot} \mathbf{V}_{CI}^2}{2} - V_{ext}\left(\overbrace{\mathbf{R}_{CI} + \varepsilon \cdot \mathbf{r}_r}^{\mathbf{r}_1}\right) + \frac{\mu \mathbf{v}_r^2}{2} - V(\mathbf{r}_r).$$

The parameter $\varepsilon'$ naturally appears in the perturbation theory in the mixed variables along with $M_1$ and $M_2$ (Chapter 2).

## 1.2. Particular Case: A Uniform External Force

A uniform force is a constant vector with no space argument:

$$M_{tot}\ddot{\mathbf{R}}_{CI} = \mathbf{F}_{ext}, \tag{9}$$

$$\mu\ddot{\mathbf{r}}_r = \mathbf{F}_r(\mathbf{r}_r) + \varepsilon \cdot \mathbf{F}_{ext}. \tag{10}$$

We first consider this case because it is simpler mathematically and is still general physically.

The first (or "particle") equation describes the centre of inertia motion with the total mass $M_{tot}$, and the internal force $\mathbf{F}_r$ is not involved there explicitly.

The second (or "wave") equation describes the relative (internal) motion (oscillations, for example) with some influence of the external force. This influence is nothing but "pumping" oscillations via action on the first particle.

In a uniform external field $\mathbf{F}_{ext} = \text{const}$ (or more generally in $\mathbf{F}_{ext} = \mathbf{f}(t)$) the two equations are totally decoupled: the two subsystems – CI and oscillator quasi-particles – do not "act" on each other and do not "see" each other as if they belonged to different non interacting physical systems. This system looks like some pointlike particle (three coordinates suffice) and electromagnetic field amplitude equations. Indeed, the "pumping" term $\varepsilon \cdot \mathbf{F}_{ext}$ in the second equation can equally be presented as due to "particle" acceleration (for a uniform force it does not mean the equation coupling though!):

$$\mu\ddot{\mathbf{r}}_r = \mathbf{F}_r(\mathbf{r}_r) + \varepsilon M_{tot}\ddot{\mathbf{R}}_{CI}(t). \tag{11}$$

This fact shows a close analogy with the electromagnetic field radiation where the source of electromagnetic waves is the charge acceleration. Then the energy $\Delta E_r$ gained by the oscillator during $0 \leq t \leq t_F$ may be called the "radiated" energy. (I speak mainly of oscillations but in fact the second equation describes the rotations too.)

An external uniform field makes two kinds of work:

1) it accelerates the CI of the whole system with mass $M_{tot}$ by "pulling" or "pushing" the first particle (first equation). The internal motion does not affect this part of work in any way. The compound nature of the whole system is taken into account only via the total mass $M_{tot}$:

$$\frac{d}{dt}\left(M_{tot}\frac{\mathbf{V}_{CI}^2}{2}\right) = \mathbf{F}_{ext} \cdot \mathbf{V}_{CI}, \tag{12}$$

2) apart from this, the external field generally "pumps" some energy into the relative motion too. This fact is contained explicitly and entirely in the *second* equation (with $\varepsilon$ as the pumping efficiency coefficient):

$$\frac{d}{dt}\left(\mu\frac{\mathbf{v}_r^2}{2} + V_r(\mathbf{r}_r)\right) = \varepsilon \cdot \mathbf{F}_{ext}\mathbf{v}_r. \tag{13}$$

These two kinds of work are additive. The energy conservation law holds and may be read now as follows: the total external field work $-\Delta V_{ext} = V_{ext}(\mathbf{r}_{1initial}) - V_{ext}(\mathbf{r}_{1final})$ is spent on the *whole system acceleration* $\Delta\left(M_{tot}\frac{\mathbf{V}_{CI}^2}{2}\right)$ and on the *internal* (relative motion) *energy increase* $\Delta E_r = \Delta\left[\mu\frac{\mathbf{v}_r^2}{2} + V_r(\mathbf{r}_r)\right]$:

$$-\Delta V_{ext} = \Delta\left(M_{tot}\frac{\mathbf{V}_{CI}^{2}}{2}\right) + \Delta E_{r}. \tag{14}$$

As we can see here, it is absolutely not necessary to insert a "radiative loss" (or "self action") term into the "particle" equation in order to respect the energy conservation law. The "particle" equation in an external field (9) should not contain "radiative losses" if it is a CI equation. The "radiated" power **taken from the external potential field** $\mathbf{F}_{ext}$ is *entirely* contained in the second (oscillator or wave) equation because it is a *relative* motion equation. Thus we can preserve the energy-momentum conservation laws in the frame of the *second order equations,* without non-physical (runaway) solutions. This is an exemplary theoretical description, and it is also valid in general case of a non-uniform external force (i.e., when $V_{ext}(\mathbf{r}_{1})$ is a non linear function of $\mathbf{r}_{1}$).

Generally, if the pumping addendum is a known function of time, the problem (11) is easy to solve exactly. We see that $\varepsilon$ here is not really a "small" coupling parameter as there is no need to take it into account perturbatively. Let us see it closer.

### *1.2.1. An Exactly Solvable 1D Problem*

Let us consider an example of exactly solvable 1D problem – an initially free harmonic oscillator ($F_{r} = -k \cdot (r_{r} - r_{r}^{eq})$) put in a uniform field $F_{ext}$ at $t = 0$. (Here $r_{r}^{eq}$ is the equilibrium distance.)

It is convenient to present the full solution at $t > 0$ in form the free initial oscillations plus the "pumped" term expressed via the retarded Green function:

$$r_{r}(t) = r_{r}^{eq} + r_{\max}\cos(\omega t + \varphi_{0}) + \int_{0}^{t}\varepsilon\frac{F_{ext}}{\mu}\frac{\sin\omega(t-t')}{\omega}dt'. \tag{15}$$

We find:

$$R_{CI}(t) = R_{CI}(0) + V_{CI}(0)\cdot t + \frac{F_{ext}}{M_{tot}}\frac{t^{2}}{2}, \tag{16}$$

$$r_{r}(t) = r_{r}^{eq} + r_{\max}\cos\left(\sqrt{\frac{k}{\mu}}\cdot t + \varphi_{0}\right) + \varepsilon\frac{F_{ext}}{k}\left[1 - \cos\left(\sqrt{\frac{k}{\mu}}\cdot t\right)\right], \tag{17}$$

$$r_{r}(0) = r_{r}^{eq} + r_{\max}\cos\varphi_{0}, \quad \dot{r}_{r}(0) = -\sqrt{\frac{k}{\mu}}\cdot r_{\max}\sin\varphi_{0}. \tag{18}$$

In the exact formulation (1)-(2) we need three measured values: $M_{1}$, $M_{2}$, and $k$. The other constants – the quasi-particle masses $M_{tot}$ and $\mu$, the dimensionless equation coupling constant $\varepsilon$, and the proper oscillator frequency $\omega_{exact} = \sqrt{k/\mu}$ – are *calculated* from these experimental data.

To find the first particle coordinate $r_{1}(t)$ we use the following relationships:

$$\underline{r_{1} = R_{CI} + \varepsilon\cdot r_{r}} = R_{CI}(0) + \dot{R}_{CI}(0)t + \frac{F_{ext}}{M_{tot}}\frac{t^{2}}{2} + \varepsilon\left\{r_{r}^{eq} + r_{\max}\cos\left(\sqrt{\frac{k}{\mu}}\cdot t + \varphi_{0}\right) + \varepsilon\frac{F_{ext}}{k}\left[1 - \cos\left(\sqrt{\frac{k}{\mu}}\cdot t\right)\right]\right\}$$

$$R_{CI}(0) = r_{1}(0) - \varepsilon\left[r_{r}^{eq} + r_{\max}\cos\varphi_{0}\right], \quad \dot{R}_{CI}(0) = \dot{r}_{1}(0) + \varepsilon\cdot r_{\max}\sqrt{\frac{k}{\mu}}\sin\varphi_{0}. \tag{19}$$

The CI coordinate (16) describes a smooth trajectory in a uniform external field. It is determined with the total mass. The relative coordinate (17) describes superposition of the *initial* and *pumped* oscillations.

And the first particle coordinate $r_1(t)$ (19) has *both* the smooth and the oscillating parts. The finite initial oscillations $\propto r_{max}\cos(\omega_{exp}t+\varphi_0), \forall t \leq 0$ are the direct classical analogue of the vacuum field fluctuation influence. On average (over many oscillation periods $\forall t$) the particle-1 coordinate is as smooth as the CI coordinate. In other words, the external force acting only on particle 1 "feels" on average the total mass $M_{tot}$ rather than $M_1$. (In case of many oscillators the particle-1 coordinate may be so fluctuating that is hardly observable as a non-averaged function of time.)

Using relations (6)-(7), we can expand the exact solutions (17) and (19) in powers of $\varepsilon'$. The corresponding series will be convergent when $|\varepsilon'|<1$ since there is nothing particular at the point $\varepsilon'=0$. These expansions are those that are obtained in the perturbation theory in the mixed variables. We are going to consider now our problem in the mixed variables since it is a formulation with a self-action which is in fact the case in CED, QED, and in the other gauge QFT.

## 2. MIXED VARIABLES

Apart from the Cartesian $(\mathbf{r}_1,\mathbf{r}_2)$ or CIRC $(\mathbf{R}_{CI},\mathbf{r}_r)$ variables, it is also possible in principle to use the first particle Cartesian ("personal") coordinates $\mathbf{r}_1$ and the relative coordinates $\mathbf{r}_r$ as a set of independent variables $(\mathbf{r}_1,\mathbf{r}_r)$:

$$\mathbf{r}_1' = \mathbf{r}_1, \quad \mathbf{r}_2' = \mathbf{r}_r = \mathbf{r}_1 - \mathbf{r}_2. \tag{20}$$

The new Lagrange function (with primes omitted) is the following:

$$L_{MIXED} = \frac{M_1 \mathbf{v}_1^2}{2} - V_{ext}(\mathbf{r}_1) + \frac{M_2 \mathbf{v}_r^2}{2} - V(\mathbf{r}_r) + M_2\left(\frac{\mathbf{v}_1^2}{2} - \mathbf{v}_1 \mathbf{v}_r\right). \tag{21}$$

The corresponding kinetic energies **are not additive** because the additive partner for the first particle energy is the second particle energy, not the internal energy. The internal energy has in its turn the CI energy as the additive partner, not the first particle energy. The term $M_2\left(\mathbf{v}_1^2/2 - \mathbf{v}_1\mathbf{v}_r\right)$ can be understood as an "interaction" Lagrangian, which bears a purely kinetic nature. The term $\propto \mathbf{v}_1\mathbf{v}_r$ in it reminds the interaction term $\propto \mathbf{j}\mathbf{A}_{rad}$ in electrodynamics. The term $M_2\mathbf{v}_1^2/2$ in it has the same meaning as the "radiative added mass" from (**I**1).

The Lagrange equations are:

$$M_1\dot{\mathbf{v}}_1 = \mathbf{F}_{ext}(\mathbf{r}_1) + M_2\left(\dot{\mathbf{v}}_r - \dot{\mathbf{v}}_1\right), \tag{22}$$

$$M_2\dot{\mathbf{v}}_r = \mathbf{F}_r(\mathbf{r}_r) + M_2\dot{\mathbf{v}}_1. \tag{23}$$

The new equations are not simpler than the initial equations (1-2). Even in absence of external field the first equation is not trivial – it has an oscillating (or more generally, "fluctuating") part $M_2\dot{\mathbf{v}}_r$. After some remarks, we will solve these *exact* equations with help of perturbation theory because such an approach is practised in CED and QED.

## 2.1. First, particle-1 equation

The form (22) is interesting with the fact that in the zeroth approximation in term $(M_2/M_1)(\dot{\mathbf{v}}_r - \dot{\mathbf{v}}_1) = \varepsilon'(\dot{\mathbf{v}}_r - \dot{\mathbf{v}}_1)$ the first equation represents the particle-1 Newton equation in the external field:

$$\dot{\mathbf{v}}_1^{(0)} = \mathbf{F}_{ext}(\mathbf{r}_1)/M_1. \tag{24}$$

It is simple, separated from the second equation, and may be thought to be good for the initial approximation. The perturbative term $\varepsilon'(\dot{\mathbf{v}}_r - \dot{\mathbf{v}}_1)$ may be understood as a "radiative loss" term, and one may be tempted to take it into account it perturbatively because, for example, in the zeroth order approximation the particle-1 equation is separated from the second one and because of smallness of the dimensionless "coupling" constant $\varepsilon' = M_2/M_1$. Let us note here that the "particle" equation (24) has a "wrong" mass $M_1$.

## 2.2. Second, "oscillator" equation

The second equation without the pumping term $M_2\dot{\mathbf{v}}_1$ describes the proper oscillator vibrations (initially free oscillations) but with some different (also "wrong") mass or frequency:

$$\dot{\mathbf{v}}_r = \mathbf{F}_r(\mathbf{r}_r)/M_2. \tag{25}$$

The right-hand term $M_2\dot{\mathbf{v}}_1$ in (23) is an oscillator energy pumping source determined with the particle-1 acceleration $\dot{\mathbf{v}}_1(t)$ at $t > 0$. When $\dot{\mathbf{v}}_1(t)$ is a known function of time, for example, $\dot{\mathbf{v}}_1^{(0)}(t)$, the equation (23) can be integrated analytically.

## 2.3. Perturbation theory in the mixed variables (Lagrange formulation)

We start from the following form of the exact equations where the formal small parameter is $\varepsilon'$:

$$\dot{\mathbf{v}}_1 = \frac{\mathbf{F}_{ext}(\mathbf{r}_1)}{M_1} + \varepsilon'(\dot{\mathbf{v}}_r - \dot{\mathbf{v}}_1), \tag{26}$$

$$\dot{\mathbf{v}}_r = \frac{\mathbf{F}_r(\mathbf{r}_r)}{M_2} + \dot{\mathbf{v}}_1. \tag{27}$$

### 2.3.1. Zeroth order 1D solutions

*Particle-1 coordinate:*

$$r_1^{(0)}(t) = r_1(0) + \dot{r}_1(0)\cdot t + \frac{F_{ext}}{M_1}\frac{t^2}{2}. \tag{28}$$

This solution is just a smooth trajectory of the first decoupled particle (i.e., with $M_1$) in the external uniform field (solution similar to elastic scattering of a charge in an external field).

*Oscillator coordinate:*

$$r_r^{(0)}(t) = r_r^{eq} + r_{max}\cos(\omega t + \varphi_0) + \varepsilon'\frac{F_{ext}}{k}(1-\cos\omega t), \quad \omega = \sqrt{\frac{k}{M_2}}. \tag{29}$$

This solution at $t>0$ is the initially "free" oscillations with the amplitude $r_{max}$ plus a "pumped" by the external force addendum. Keeping the latter in the oscillator zeroth-order solution $r_r^{(0)}(t)$ is analogous to the bremsstrahlung calculation in CED and in certain cases of QED: first we solve the "mechanical" part of the problem, obtain $r_1^{(0)}(t)$, and then we solve the wave equation with the known source $\dot{\mathbf{v}}_1^{(0)}(t)$.

Let us repeat that the oscillator frequency $\omega = \sqrt{k/M_2}$ in this approximation is somewhat different from the exact proper frequency $\omega_{exact} = \sqrt{k/\mu}$.

### 2.3.2. First Order 1D Solutions

*Particle-1 coordinate:*

Formally we obtain the following solution in the first order:

$$r_1^{(1)}(t) = \left[r_1(0) - \varepsilon'\left(r_r^{eq} + r_{max}\cos\varphi_0\right)\right] + \left[\dot{r}_1(0) + \varepsilon'\omega r_{max}\sin\varphi_0\right]\cdot t + \frac{F_{ext}(1-\varepsilon')}{M_1}\frac{t^2}{2} + $$
$$+\varepsilon'\left[r_r^{eq} + r_{max}\cos(\omega t + \varphi_0) + \varepsilon'\frac{F_{ext}}{k}(1-\cos\omega t)\right] \tag{30}$$

The particle-1 solution obtained a correction to the smooth part of the trajectory (the acceleration term is decreased on average) and some oscillating addenda due to connecting to the particle-2 with help of a spring. The oscillating in time addendum remains oscillating even if there is no external force at all. The constant and the linear in time corrections are grouped with the particle-1 initial position and velocity. They cancel the corresponding contribution of the oscillating term to the initial data.

*Oscillator coordinate:*

$$r_r^{(1)} = r_r^{eq} + r_{max}\left[\cos(\omega t + \varphi_0) - \frac{1}{2}\varepsilon'\omega t\cdot\sin(\omega t + \varphi_0)\right] + \varepsilon'(1-\varepsilon')\frac{F_{ext}}{k}\left[1 - \left(\cos\omega t - \frac{1}{2}\varepsilon'\omega t\cdot\sin\omega t\right)\right] \tag{31}$$

The oscillator solution obtained some time-dependent corrections that are supposed to "tune" its initially inexact frequency $\omega = \sqrt{k/M_2}$ closer to $\omega_{exact} = \sqrt{k/\mu}$. As well, the pumping force obtained an additional factor $(1-\varepsilon')$.

### 2.4. Analysis and Improvement of the Analytical Solutions

*First order particle solution*

First of all we see that the acceleration part $\frac{F_{ext}}{M_1}\frac{t^2}{2}$ of $r_1^{(1)}(t)$ obtained a correction represented as the factor $(1-\varepsilon')$ at the external force. It seems the first particle is accelerated now *less* than in the zeroth order - as if it had now a heavier mass $M_1/(1-\varepsilon')$.

What is it - the mass $M_1$ "numerical or theoretical correction"? Should the mass $M_1$ acquire theoretical corrections? The right answer depends completely on our understanding of the coupling consequence and on the mass measurement procedure. If our bound system is literally assembled from two experimentally separable and measured masses $M_1$ and $M_2$ by connecting them with a massless spring with a known $k$ (a kit-made system), then these perturbative corrections are numerically meaningful. Of course, it is not the particle-1 mass correction; it is the particle-1 *solution correction*, i.e. taking into account its binding to the other system parts. *On time average* the first *bound* particle "looks" indeed heavier than the non-bound one. Hence, "switching on" the "radiative loss" term $\varepsilon'(\dot{\mathbf{v}}_r - \dot{\mathbf{v}}_1)$ for particle 1 $\mathbf{F}_{ext}(\mathbf{r}_1)/M_1 \to \mathbf{F}_{ext}(\mathbf{r}_1)/M_1 + \varepsilon'(\dot{\mathbf{v}}_r - \dot{\mathbf{v}}_1)$ means, on one hand, a smaller *average* acceleration in an external field. In this connection let us introduce in our theoretical description an "effective inverse mass"

$$1/\left(M_1^{(1)}\right)_{eff} = (1-\varepsilon')(1/M_1), \tag{32}$$

which is numerically closer to $1/M_{tot}$ and thus the smooth part of solution $r_1^{(1)}$ is also closer to $r_1$ than $r_1^{(0)}$.

Next, if the external force starts acting when the oscillator passes its equilibrium position $F_r = 0$ (it is possible at $\varphi_0 = \pi/2 + \pi n$ or when $r_{max} = 0$), then during the time interval much shorter than the proper oscillation period ($\omega t_F \ll 1$) the external force only slightly displaces the first particle. The positive pumped term in the first-order solution $\left(\frac{M_2}{M_1}\right)^2 \frac{F_{ext}}{k}(1-\cos\omega t) \approx \left(\frac{M_2}{M_1}\right)^2 \frac{F_{ext}}{k}\frac{\omega^2 t^2}{2} = \frac{M_2}{M_1}\frac{F_{ext}}{M_1}\frac{t^2}{2}$ cancels the negative "mass correction" contribution. As a result, the external force during such an action "feels" only non-bound first particle with the true $M_1$ (as if it were *free*). It is also physically correct picture. (This fact can be used for measuring the pure $M_1$ with a *pulse* force in a compound system.)

In case of such a short push ($F_r = 0$, $\omega t_F \ll 1$) the first particle obtains some initial velocity $|\mathbf{v}_1|_{init}$ which decreases *after* the force stops acting. Let us note that this decrease happens not because of "friction" or "self-action" but due to transmitting some energy into the potential energy of the relative motion. In fact the velocity $\mathbf{v}_1(t)$ becomes oscillating with some average value $|\langle\mathbf{v}_1\rangle| = |\mathbf{V}_{CI}|$, which is, of course, smaller than $|\mathbf{v}_1|_{init}$.

The oscillating addendum in $r_1^{(1)}(t)$ describes influence of the system internal interaction on the particle-1 coordinate. It appears starting from the first order of the perturbation theory, even in absence of external force. Hence, "switching on" the "radiative loss" for particle 1 means, on the other hand, introducing fluctuations of its coordinates. If we are conscious of what a system we describe, then this agrees with our understanding of the coupling effect and it is justified numerically if $\varepsilon' \ll 1$. The first order solution $r_1^{(1)}(t)$ is therefore correct from the physical and mathematical viewpoints: we see that it is closer to the exact solution.

*First order relative motion solution*

The PT-corrections to the relative coordinate improve the solution "frequency behavior" even in absence of external force. It is normal since in the zeroth-order of PT the frequency $\omega$ is not exact – it is calculated via $M_2$ rather than via $\mu$. We note however that the linearly diverging in time amplitude worsens the numerical accuracy of the first order solution $r_r^{(1)}(t)$ at big times $\omega t_F \gg 1$.

Upon understanding all that, we can make the first order solution $r_r^{(1)}(t)$ more compact analytically and more accurate numerically if we represent some of PT corrections as "corrections" to $M_2$. Namely, the first order correction to both cosines may be represented as originating from a numerical factor at $M_2$ in their arguments: $M_2(1-\varepsilon')$. Now let us introduce $(M_2)_{eff}$ in a similar way:

$$\cos\omega t - t\frac{1}{2}\varepsilon'\cdot\omega\cdot\sin\omega t \approx \cos\sqrt{\frac{k}{M_2(1-\varepsilon')}}\cdot t \;, \quad (M_2)_{eff}^{(1)} = (1-\varepsilon')M_2, \tag{33a}$$

or in terms of frequency:

$$\sqrt{\frac{k}{(M_2)_{eff}^{(1)}}} = \omega_{eff}^{(1)} = \omega/\sqrt{1-\varepsilon'}. \tag{33b}$$

With these definitions the solution $r_r^{(1)}(t)$ does not diverge with time anymore:

$$r_r^{(1)} \approx r_r^{eq} + r_{max}\cos\left(\omega_{eff}^{(1)}t + \varphi_0\right) + \varepsilon'(1-\varepsilon')\frac{F_{ext}}{k}\left(1 - \cos\omega_{eff}^{(1)}t\right). \tag{34a}$$

Of course, (33a) is not the mass $M_2$ numerical correction; it is the *oscillator solution (frequency) correction* that can be presented in this formula as $M_2$ correction. In the first order such a solution is closer to the exact one. If we did not know the origin of the equations (26)-(27) and their exact solutions, then introducing $(M_2)_{eff}^{(1)}$ or $\omega_{eff}^{(1)}$ might be considered as a kind of Bogolubov-Mitropolsky approach to solving differential equations – developing the oscillator phase in powers of small parameter $\varepsilon'$. Such a technique extends the good accuracy of asymptotic solutions to the region of big times $\omega t_F \gg 1$.

The pumping addendum to the free initial oscillations in formula (34a) (the term proportional to $F_{exp}$) is the oscillator excitation due to the first particle acceleration. This is a desired result. However in the first order it is proportional to the polynomial $\varepsilon'(1-\varepsilon')$. So we may denote the latter as an effective "coupling" constant:

$$\left(\varepsilon'_{eff}\right)^{(1)} = \varepsilon'(1-\varepsilon'). \tag{35}$$

Then we obtain the effective first order solution $r_r^{(1)}{}_{eff}$ in the following short form:

$$r_r^{(1)}{}_{eff} = r_r^{eq} + r_{max}\cos\left(\omega_{eff}^{(1)}t + \varphi_0\right) + \varepsilon'_{eff}{}^{(1)}\frac{F_{ext}}{k}\left(1 - \cos\omega_{eff}^{(1)}t\right). \tag{34b}$$

With $\omega_{eff}$, $(M_2)_{eff}$, and $\varepsilon'_{eff}$ the first order solution $r_r^{(1)}{}_{eff}$ (34b) is compact and is much better defined for **big** times $\omega t_F \geq 1$ than (31).

### 2.5. Effective Solutions in the n-th perturbative order

In the n-th PT order we may as well reduce the long PT series to short formulas with help of "hiding" (summing) the powers of $\varepsilon'$ into effective parameters $(M_1^{-1})_{eff}^{(n)}$, $\omega_{eff}^{(n)}$ or $(M_2)_{eff}^{(n)}$, and $\left(\varepsilon'_{eff}\right)^{(n)}$. In particular, the coupling constant $\varepsilon'$ at $r_r^{(1)}$ in $r_1^{(1)}$ (30) will also be "transformed" into $\varepsilon'_{eff}$ in higher than the first PT orders. This statement is easily proved with comparing the exact solutions (17)-(19) with the effective ones: $r_1 = R_{CI}\left(M_{tot}(M_1, \varepsilon'), \mu(M_2, \varepsilon'), \varepsilon(\varepsilon')\right) + \varepsilon(\varepsilon') \cdot r_r\left(\mu(M_2, \varepsilon'), \varepsilon(\varepsilon')\right)$, $r_r = r_r\left(\mu(M_2, \varepsilon'), \varepsilon(\varepsilon')\right)$.

Let us denote the set of exact fundamental (quasi-particle) constants as $C_{exact}(\varepsilon')$ and the approximate (effective) ones as $C_{eff}^{(n)}(\varepsilon')$ for short.

When the numerical values of $M_1$ and $M_2$ are such that $\varepsilon' \ll 1$, the PT series of the effective constants $C_{eff}^{(n)}$ (polynomials in powers of $\varepsilon'$) converge quickly to $C_{exact}(\varepsilon')$ and in general case of kit-made systems the approximate solutions $\left(r_1^{(n)}\right)_{eff} = r_1\left(C_{eff}^{(n)}\right)$ and $\left(r_r^{(n)}\right)_{eff} = r_r\left(C_{eff}^{(n)}\right)$ converge to the exact ones $r_1(t)$ and $r_r(t)$ with $n \to \infty$.

The numerical "corrections" to the "bare fundamental constants" *in approximate solutions* (29), (30) make sense since in course of perturbative calculations the "bare" fundamental constants transform from initially inexact (but known) *particle* values $1/M_1$, $M_2$, and $\varepsilon'$ into the exact ("dressed" or calculated) *quasi particle* ones (7) in the solutions $\left(r_1^{(n)}\right)_{eff}$ and $\left(r_r^{(n)}\right)_{eff}$. The latter *functionally* coincide with the exact solutions for $n \geq 1$. As the functional forms are already correct $\left(r^{(n)}\right)_{eff} = r(C_{eff}^{(n)})$, solely numerical values of inexact constants $C_{eff}^{(n)}$ perturbatively change (converge) in them when $n \to \infty$. (This simplicity is attained only in a uniform external force.)

*Case of big $\varepsilon'$*

If $\varepsilon' \geq 1$, the perturbative series for effective constants diverge and we must apply a *non-linear summation* of the mass and charge "corrections". The simplest non-linear positive Padé approximant $1 - \varepsilon' + \varepsilon'^2 + ... \approx 1/(1+\varepsilon')$ turns out numerically to be the best in our problem. Then our effective constants just coincide with the exact ones (7) in all higher orders and the effective solutions coincide with the exact ones (17)-(19).

Although demonstrated on a 1D case, this approach (i.e., introducing the effective constants) is valid in the general 3D case too. So, if the system is assembled from known parts we can build reasonable perturbative solutions in the mixed variables.

### 2.6. Why the fundamental constants obtain "corrections"?

How to explain this remarkable property of the solutions that permits to radically simplify their analytical expressions with "hiding" (summing) all perturbative corrections into the effective constants? The answer clue is in the kinetic nature of the perturbative terms - they are proportional to accelerations with some mass coefficients in the mixed variable formulation. That is why their corrections contribute effectively and (starting from the first order) solely into the masses and into the "coupling" constant as some scaling factors. Indeed, let us remember equation (I1). It is one equation with one kinetic "perturbative" term, so only *one* constant $m_e$ obtains "corrections" if the term $\delta m \cdot \ddot{\mathbf{r}}$ considered perturbatively. The formulation (26)-(27) consists of *two* coupled equations with kinetic perturbations so *two* "initially incorrect" masses obtain perturbative corrections. In addition, the coupling parameter obtains "corrections" too since it is also determined with the correct masses. The formulation (26)-(27) is equivalent to the CIRC formulation (9)-(10) with the relationship (8). With the same success we can develop the perturbation theory in the system (6)-(10) casted in the form:

$$\mathbf{r}_1 = \mathbf{R}_{CI} + \varepsilon \cdot \mathbf{r}_r, \qquad \ddot{\mathbf{R}}_{CI} = \mathbf{F}_{ext}/M_{tot}, \qquad \dot{\mathbf{v}}_r = \mathbf{F}_r(\mathbf{r}_r)/\mu + (\varepsilon/\mu) \cdot (\dot{\mathbf{v}}_1 - \varepsilon \dot{\mathbf{v}}_r),$$

$$1/M_{tot} = 1/[M_1(1+\varepsilon')], \quad 1/\mu = (1+\varepsilon')/M_2, \quad \varepsilon = \varepsilon'/(1+\varepsilon').$$

From here we clearly see that not only two "bare" masses obtain corrections when the system is solved perturbatively but also the equation coupling constant $\varepsilon'$ to be finally expressed via the exact ones.

In case of kit-made systems, taking them into account perturbatively is possible but not really necessary since we may rewrite the original exact equations (22), (23) in a way to include exactly the kinetic part of interaction into the zeroth-order approximation from the very beginning. We may start from (3), (7), and (8) that contain the exact expressions of particle coordinate via quasi-particle ones. Then we transform the equations in a simple way:

$$M_2 \dot{\mathbf{v}}_r = \mathbf{F}_r(\mathbf{r}_r) + M_2 \dot{\mathbf{v}}_1 \Rightarrow$$

$$M_2 \left( \dot{\mathbf{v}}_r - \underbrace{\frac{M_2}{M_1+M_2} \dot{\mathbf{v}}_r}_{added\,to\,the\,left-hand\,side} \right) = \mathbf{F}_r(\mathbf{r}_r) + M_2 \left( \dot{\mathbf{v}}_1 - \overbrace{\underbrace{\frac{M_2}{M_1+M_2} \dot{\mathbf{v}}_r}_{added\,to\,the\,right-hand\,side}}^{\ddot{\mathbf{R}}_{CI}} \right) \Rightarrow \mu \dot{\mathbf{v}}_r = \mathbf{F}_r(\mathbf{r}_r) + M_2 \dot{\mathbf{V}}_{CI} . \quad (36)$$

The zeroth order equation for $\mathbf{v}_r$ (23) (where $\dot{\mathbf{v}}_1$ is a known function of time expressed via $F_{ext}$) coincides in functional form with the exact one (36). So the zeroth order solution differs from the exact one only with the numerical value of the reduced mass in the oscillator frequency and a coefficient at the external force (pumping efficiency coefficient). The corresponding numerical factors, as we could see, can be represented as series in powers of $\varepsilon'$ ("corrections" to the "bare" values).

$$M_1 \dot{\mathbf{v}}_1 = \mathbf{F}_{ext}(\mathbf{r}_1) + \underbrace{M_2(\dot{\mathbf{v}}_r - \dot{\mathbf{v}}_1)}_{\leftarrow move\,to\,the\,left\,hand\,side} \Rightarrow \underbrace{(M_1+M_2)}_{M_{tot}} \underbrace{\left( \dot{\mathbf{v}}_1 - \frac{M_2}{M_1+M_2} \dot{\mathbf{v}}_r \right)}_{\ddot{\mathbf{R}}_{CI}} = \mathbf{F}_{ext} \Rightarrow M_{tot} \ddot{\mathbf{R}}_{CI} = \mathbf{F}_{ext}(\mathbf{r}_1) . \quad (37)$$

The zeroth order equation for $\mathbf{r}_1$ (24) coincides in functional form with the exact one for $\mathbf{R}_{CI}$ (36). The relation (8) is linear in the variables $\mathbf{R}_{CI}$ and $\varepsilon \cdot \mathbf{r}_r$ whose equations are determined with quasi particle constants. The n-th order solution $r_1^{(n)}(t)$ obtained with PT is factually expansion of the relations (7)-(8) in powers of $\varepsilon'$: $\mathbf{r}_1 = \mathbf{R}_{CI}\left(M_{tot}(M_1,\varepsilon'),\mu(M_2,\varepsilon'),\varepsilon(\varepsilon')\right) + \varepsilon(\varepsilon') \cdot \mathbf{r}_r\left(\mu(M_2,\varepsilon'),\varepsilon(\varepsilon')\right)$.

Now, if the external force is not uniform $\mathbf{F}_{ext}(\mathbf{R}_{CI}+\varepsilon\mathbf{r},t) \neq const$, it can be developed in series "around" $\mathbf{R}_{CI}$ and thus give some perturbative terms in the frame of CIRC formulation. These terms will not contribute to the quasi particle constants and the analytical expressions will not reduce to short formulas. This is a technically and physically correct and transparent approach to the kit-made system description.

Generalizing we can say that a particle-like equation with an external field solely may be the CI equation of a compound system with inexact argument – $\mathbf{R}_{CI}$ instead of the exact $\mathbf{R}_{CI}+\varepsilon \cdot \mathbf{r}_r$. And an oscillator-like equation with a pumping term $\propto \ddot{\mathbf{R}}_{CI}$ may be the relative (internal) motion equation in a compound system description. This possibility indicates how the exact description could be "restored" from classical equations – by restoring the right argument in the external force and by the passage to the right meanings of variables. It will come in handy in the next chapters. Finishing this one, let us underline that not all compound systems in nature are made of a kit (mountable-dismountable). There are compound systems that cannot be dismounted in principle. First of all, this is the case of charges and electromagnetic waves. The chapter describing their "interaction" is not finished yet. We will try to reveal the dead-end with it considering a "welded" mechanical system.

### 3. THEORY CONSTRUCTION FOR A WELDED (NON MOUNTABLE) COMPOUND SYSTEM

Let us imagine now a possible historical development of "theoretical physics" when the two bound bodies cannot be ever separated in reality, like in Fig. 2. They are permanently bound by definition. This case is radically different from the kit-made system because the measured experimentally fundamental constants are quite different. We make on purpose our consideration somewhat similar to the historical development of electrodynamics to better understand the evolution of our physical notions (including renormalizations).

Firstly, the visible body is small itself, studied from relatively big distances and then, due to relatively small amount of energy normally stored in the oscillation motion, we first do not really know that our particle-1 (shell) is not "elementary". We measure the system weight with a balance and we think that this way gives just the body or particle-1 mass $M_1$. This stage corresponds to the Classical Mechanics: we write the Newton equation without "losses" and sincerely think that *three* coordinates $\mathbf{r}_1(t)$ are *sufficient* to describe its motion. According to the experimental methodology (we do not hurry, the measurement time is much longer than the oscillation period), we obtain the total mass as the experimental mass value for our solution - $(M_1)_{\exp} = M_{tot}$. It gives a good average prediction of the first particle (shell) trajectory in an external field. Hence, for a welded system we obtain:

$$\ddot{r}_1(t) = \frac{F_{ext}}{(M_1)_{\exp}}, \qquad r_1(t) = r_1(0) + \dot{r}_1(0) \cdot t + \frac{F_{ext}}{(M_1)_{\exp}} \frac{t^2}{2}. \tag{38}$$

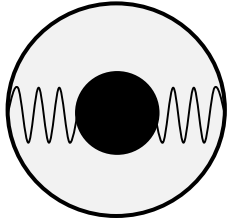

Fig. 2. Mechanical non-elementary system with a hidden 1D oscillator.

Thus the pointlike particle notion finds a good experimental support and this is what the Classical Mechanics is about.

Much later we discover experimentally that the body *acceleration* excites some waves observed indirectly at a distance as some (ultra) sound (or light waves) of a certain frequency $\omega_{\exp}$. This frequency is the second (after mass) fundamental constant of our body (or better, of our physical system). If the wave decrement is small, then the frequency $\omega_{\exp}$ is a well-defined number (a sharp line).

About the waves we sincerely think that they may exist and exist quite separately (independently) from our body. To describe them phenomenologically we try an oscillator equation with a pumping term expressed via the particle-1 acceleration, for example:

$$\ddot{\mathbf{r}}_{osc} + (\omega_{\exp})^2 \mathbf{r}_{osc} = (\alpha_p)_{\exp} \dot{\mathbf{v}}_1. \tag{39}$$

The true oscillator amplitude $\mathbf{r}_{osc}$ may be unknown to us at the beginning but we suppose that the observed *sound* (or light wave) amplitude is simply proportional to the true $\mathbf{r}_{osc}(t)$.

The oscillator proper frequency $\omega_{osc}$ is already known experimentally, so it is a measured rather than a calculated value. We ask experimentalists to measure everything else what is possible and meanwhile we are trying to build a theoretical description of this phenomenon. For the time being we consider the pumping efficiency or the equation "coupling constant" $(\alpha_p)_{\exp}$ to be known experimentally too, at least for the observable sound (or light) wave amplitudes. In particular, in the 1D solution with a uniform external force it 4stands as an experimental coefficient $(\alpha_p)_{\exp}$ at $F_{ext}$ (in order to simplify the solutions we assume that $r_{\max} = r_r^{eq} = 0$):

$$r_{osc} = \frac{(\alpha_p)_{\exp}}{(\omega_{\exp})^2} \frac{F_{ext}}{(M_1)_{\exp}} \cdot (1 - \cos \omega_{\exp} t), \ 0 \leq t \leq t_F, \quad (40a)$$

$$r_{osc} = r_{\max} \cos(\omega_{\exp} t + \varphi_F), \ t \geq t_F. \quad (40b)$$

Let us admit that the description (40) also agrees nicely with numerous experimental data: the sound (or light) amplitude is linear in $F_{ext}$ and in $(1 - \cos \omega_{\exp} t)$ when $0 \leq t \leq t_F$, and in $(1 - \cos \omega_{\exp} t_F)$ when $t \geq t_F$, so $|(\alpha_p)_{\exp}| > 0$. We are happy with the accuracy of our phenomenological description of waves. This stage of our theory development (i.e. (39)) corresponds roughly to the Maxwell theory of EM wave radiation due to charge acceleration (a wave equation with a known source, Larmor formulas).

### 3.1. A "self-consistent" theory with a "self-action"

According to the wave equation (39) and the solution (40) the wave acquires and carries away (or accumulates) some energy-momentum. To calculate it properly, we must know the true oscillator amplitude, the mass $M_{osc}$ and/or elasticity $k$. We can't wait for the corresponding experimental data. But anyway, in our present notions, this energy-momentum is evidently "taken" from the particle-1 energy-momentum: the wave source is determined with the body (particle-1) acceleration. For this reason we want also to consider the radiation damping ("friction") effect in the particle-1 equation for the sake of the energy-momentum *conservation*, as it should be according to our previous experience with the usual *mechanical* friction. We believe that our body is elementary (pointlike) and therefore there should be *some additional force* (friction-like) term in the particle equation responsible for the radiative losses. We want to build a self-consistent theory from the "first principles" well established and justified up to now, i.e. from the Classical Mechanics principles with its "top" achievement – the variational principle.

The particle equation (38) and the wave equations (39) without pumping term can be obtained from the corresponding Lagrangians. Now we construct an "interaction" Lagrangian with the bilinear product term $-(\alpha_p)_{\exp} M_{osc} \mathbf{v}_1 \cdot \mathbf{v}_{osc}$ to obtain the right pumping term in (39) from the variational principle. Being experienced and clever theorists, we admit also the existence of a quadratic "loss" term $\alpha_{loss} \cdot M_{osc} \mathbf{v}_1^2 / 2$ that does not make a direct contribution to the oscillator equation but may contribute to the particle-1 equation. So our full "self-consistent" Lagrangian may have in general case the following form:

$$L_{trial} = \frac{(M_1)_{\exp} \mathbf{v}_1^2}{2} - V_{ext}(\mathbf{r}_1) + \frac{M_{osc} \mathbf{v}_{osc}^2}{2} - \frac{k \cdot \mathbf{r}_{osc}^2}{2} + M_{osc}\left(\alpha_{loss} \frac{\mathbf{v}_1^2}{2} - (\alpha_p)_{\exp} \mathbf{v}_1 \mathbf{v}_{osc}\right). \quad (41)$$

(We easily recognize in (41) a mixed variable Lagrangian.) The corresponding equations are:

$$\dot{\mathbf{v}}_{osc} + (\omega_{\exp})^2 \mathbf{r}_{osc} = (\alpha_p)_{\exp} \dot{\mathbf{v}}_1(t). \quad (42)$$

$$\dot{\mathbf{v}}_1 = \frac{\mathbf{F}_{ext}(\mathbf{r}_1)}{(M_1)_{\exp}} + \frac{M_{osc}}{(M_1)_{\exp}}\left[(\alpha_p)_{\exp} \dot{\mathbf{v}}_{osc} - \alpha_{loss} \dot{\mathbf{v}}_1\right]. \quad (43)$$

arXiv:0811.4416

The oscillator equation has seemingly not changed. The particle equation has acquired some "radiative loss" terms (kind of "self-action").

If the "radiative loss" terms are treated perturbatively (with the small parameter $M_{osc}/(M_1)_{exp}$), then our solutions (38) and (40) become the zeroth-order approximations. We could go now to considering the higher order corrections,(as previously, then introduce the effective constants, etc., but we will not do it. Instead we rewrite the exact equations of our "self-consistent" system (41)-(43) in the form of CIRC variables (36), (37):

$$\dot{\mathbf{v}}_{osc} + \frac{(\omega_{exp})^2}{1-(\alpha_p)_{exp}\beta}\mathbf{r}_{osc} = \frac{(\alpha_p)_{exp}}{1-(\alpha_p)_{exp}\beta}(\dot{\mathbf{v}}_1 - \beta\dot{\mathbf{v}}_{osc}), \qquad (44)$$

$$\left[(M_1)_{exp} + \alpha_{loss}M_{osc}\right]\left[\dot{\mathbf{v}}_1 - \frac{M_{osc}(\alpha_p)_{exp}}{\left[(M_1)_{exp} + \alpha_{loss}M_{osc}\right]}\dot{\mathbf{v}}_{osc}\right] = \mathbf{F}_{ext}(\mathbf{r}_1). \qquad (45)$$

With choosing $\beta = \dfrac{M_{osc}(\alpha_p)_{exp}}{(M_1)_{exp} + \alpha_{loss}M_{osc}}$ and denoting $\dot{\mathbf{v}}_1 - \beta\dot{\mathbf{v}}_{osc} = \dot{\mathbf{V}}$ we obtain the CIRC-like equations and the corresponding *exact* solutions $\mathbf{r}_{osc}(t)$ and $\mathbf{r}_1(t)$ when $\mathbf{F}_{ext} = const$ with the following "effective" constants:

$$\omega_{eff} = \omega_{exp}/\sqrt{1-(\alpha_p)_{exp}\beta}, \quad (M_1)_{eff} = (M_1)_{exp} + \alpha_{loss}M_{osc}, \quad (\alpha_p)_{eff} = (\alpha_p)_{exp}/\left[1-(\alpha_p)_{exp}\beta\right].$$

Using these effective constants in the exact solutions worsens seriously the agreement with experiments: the oscillator experimental frequency $\omega_{exp}$ is different from $\omega_{eff}$ appeared in (44). The smooth part of the particle-1 trajectory is determined with the value of $(M_1)_{exp}$ which is different from $(M_1)_{eff}$, etc. What a bad surprise! We see that our attempt of taking into account the energy-momentum losses of the first particle in a "self-consistent" way (41) fails: it leads to disagreement with the experimental data which are still so fine described with the "zeroth" approximations (38) and (40). Finite or infinite, the "theoretical" corrections to the *experimental* constants are not necessary at all. If correction to $(M_1)_{exp}$ can be removed by choosing $\alpha_{loss} = 0$, this simple way does not work for corrections to the oscillator frequency (or the oscillator mass $M_{osc}$) since $(\alpha_p)_{exp} \neq 0$ in (39), (40). If our experimentalists came now with the measured values of $M_{osc}$, $k$, and $(\alpha_p)_{exp}$, we would not be able put them in use in our "self-consistent" theory!

The solutions of the system (44), (45) (i.e., with the effective constants) may be called the *non renormalized* exact solutions of the "self-consistent" theory (41).

But why does the oscillator frequency change in transition from (39) to (42)? Our goal was just to take into account the radiative losses in the *particle* equation, not to intervene into the *oscillator* one. The answer is that although the oscillator equation has not changed its form, *its coupling* to the modified particle equation leads to the frequency change. The pumping term $(\alpha_p)_{exp}\dot{\mathbf{v}}_1$, being in (39) a known function of time $\propto \mathbf{F}_{ext}(t)$ (just an external source, the equations decoupled), has become an *unknown variable* in (42). It is so since the loss term added to the right-hand side of the particle equation $\propto \dot{\mathbf{v}}_{osc}$ is a too radical intervention in the mathematical description – it is of the *second* order in the time derivative $d/dt$. It is not "small" but as "big" as the main kinetic term $M_{osc}\dot{\mathbf{v}}_{osc}$. The same statement is valid for the first particle equation. Casting the system (42), (43) in the CIRC form (44), (45) makes it evident.

This situation is similar to the Lorentz trial equation (**I**2) where the "electromagnetic mass" term worsens the results.

Hence we must recognize this way of the theory "adjusting" (41)-(43) rebuilds unexpectedly the equations, brings up contradictions with the experimental data and therefore cannot be accepted. Similar (perturbative) corrections are "obtained" by the fundamental constants in QED. So the same conclusion is valid for the standard QED perturbative and exact *non-renormalized* solutions – they are non-physical on the same ground. There must be another way of the "radiative loss" taking into account, without kinetic perturbative terms (without self-action).

Do we have a hint about this "another" way? Yes, we do. As far as the electron is permanently "coupled" to the quantized electromagnetic field, they both represent a compound system with its centre of inertia and relative coordinates, and a CIRC approach is the only right way to describe it. This approach is free from logical and mathematical difficulties and it describes the experimental data with a perfect precision. We will develop it in the next chapters and now we will explain what the famous renormalizations are about.

### *3.1.1. Renormalizations*

In QED, after many years of stalling, the (infinite in value) corrections to the fundamental constants are *discarded*. It has become a common practice in the renormalizable theories. It means *postulating* in each PT order new equations, just as in (**I**3). As I said above, finite or infinite, the "theoretical" corrections to the initial – *experimental* – constants are not necessary at all. In particular, the term $\propto \alpha_{loss}\dot{\mathbf{v}}_1$ in the "particle" equation (43) is of the same meaning as the "electromagnetic mass" from (**I**2) – it is good for nothing.

Of course, such a discarding is not justified mathematically. It is nonsense. The right conclusion is that the "self-consistent" theory (in the Lorentz spirit = self-action)) is not good to describe the experimental data.

We know well how the physicists "justify" the correction discarding: they call it the constant renormalizations. Let us also discard our "corrections" to the fundamental constants in the exact *equations* (44) and (45) on the same "ground". In our case of exact solutions it means their exact (rather than perturbative) constant renormalizations:

$$\omega_{eff} = \omega_{\exp}/\sqrt{1-\cancel{(\alpha_p)_{\exp}\beta}}, \qquad (\alpha_p)_{eff} = (\alpha_p)_{\exp}/\left[1-\cancel{(\alpha_p)_{\exp}\beta}\right],$$
$$(M_1)_{eff} = (M_1)_{\exp} + \cancel{\alpha_{loss}M_{osc}}, \quad \Rightarrow \beta = \frac{M_{osc}}{(M_1)_{\exp}}(\alpha_p)_{\exp}. \qquad (46)$$

Then we discover, with even bigger surprise, that the so violently obtained "renormalized" equations are as good as or better than the zeroth order ones. Indeed, denoting their exact (exactly "renormalized") solutions with the tildes on the top, we see that:

$$\tilde{r}_{osc}(t) = \frac{(\alpha_p)_{\exp}}{(\omega_{\exp})^2}\frac{F_{ext}}{(M_1)_{\exp}}(1-\cos\omega_{\exp}t), \qquad (47)$$

$$\tilde{r}_1(t) = r_1(0) + \frac{F_{ext}}{(M_1)_{\exp}}\frac{t^2}{2} + \left[(\alpha_p)_{\exp}\frac{M_{osc}}{(M_1)_{\exp}}\right]^2\frac{F_{ext}}{k}(1-\cos\omega_{\exp}t). \qquad (48)$$

The oscillator solution has not changed, so it is as perfect as the original phenomenological solution (40). The particle "renormalized solution" has preserved its smooth part and has obtained an oscillating term. It has become better if compared to more precise experimental data. In fact, the curve (48) coincides with the exact one if we remember that in our welded system $M_{osc} = \mu$, $(M_1)_{exp} = M_{tot}$, and $(\alpha_p)_{exp} = \varepsilon M_{tot}/\mu$. At this moment we may suppose that the experimentalists bring fresh and improved experimental curves of $r_1(t)$ obtained with high-resolution high-speed film camera where the oscillations of $r_1(t)$ are *observable*. Our good theoretical predictions based on the constant "renormalizations" look as a miracle! Isn't it a triumph of renormalizations?!

Let us not fool ourselves. The theory "development" – passage from (38)-(40) to (41)-(43) is physically wrong – it does not describe the experimental data unlike the original equations (38)-(40). This theory is based on the self-action idea: introducing additional kinetic terms to the right-hand side of the particle equation for the sake of respecting the conservation laws. The "self-consistent" solutions obtain corrections to the fundamental constants and this is the *main* reason of their *deviations* from the experimental data. It is already a sufficient reason to abandon this way of the theory constructing. But we "continue" to deal with it with help of "doctoring" the constant corrections. The mathematically inacceptable discarding these corrections (= the renormalization prescription) restores the right constants. The rest depends on luck. Indeed, apart from constant "corrections", the first particle solution acquires an oscillating term $\varepsilon^2 \frac{F_{ext}}{k}(1-\cos\omega_{exp}t)$ that remains after renormalizations. With correct constants in it this term describes the correct physics: the internal motion influence on the particle-1 coordinate. (In our particular case with $r_r^{eq} = r_{max} = 0$ it is the sole term originating from $\varepsilon \cdot \mathbf{r}_r$ in the first and higher PT orders.) Obtained perturbatively or exactly, this term is necessary, as we have made sure in the CIRC- and mixed-variable formulation (20)-(23).

As the mixed-variable formulation (20)-(23) for a kit-made system in the perturbative approach departs from inexact ("bare") constants, the perturbative corrections to them in the *solutions* are meaningful. But the "self-consistent" (and accidentally mixed-variable) formulation (41)-(43) for a welded system in the perturbative approach departs from the *right* ("dressed" or quasi-particle) constants, so the perturbative corrections to them are *meaningless*. On the other hand, the mixed variable formulation gives the oscillating term too. That is why the renormalizations "work": they discard the unnecessary addenda to the experimental constants but leave intact the correct oscillating term $\varepsilon^2 \frac{F_{ext}}{k}(1-\cos\omega_{exp}t)$ appearing in the first PT order (or in the exact solution). It is a huge luck. We say that our theory is "renormalizable". But there are non-renormalizable theories too (the brightest example is the quantum gravity).

Many theorists recognize the mathematical weakness of renormalizations and thus the physical inconsistency of the theory as the main reason of mathematical difficulties. But most resort to the Wilson's "interpretation". As it does not help in the non-renormalized theories, a finite "fundamental" distance was introduced to have a natural cut-off (string and superstring theories).

Let us underline again – in case of a "welded" system we start from the experimental (rather than from "bare") constants in (41)-(43) because they were introduced and measured at the *previous stages* of the physics development:

1)  $\ddot{\mathbf{r}}_1(t) = \dfrac{\mathbf{F}_{ext}}{(M_1)_{exp}}$  - Newton mechanics,

2)  $\ddot{\mathbf{r}}_{osc} + (\omega_{exp})^2 \mathbf{r}_{osc} = (\alpha_p)_{exp} \dot{\mathbf{v}}_1$  - Maxwell electrodynamics (EM wave radiation with known $\dot{\mathbf{v}}_1$).

Our trial Lagrangian for a "self-consistent" theory

$$L_{trial} = \frac{(M_1)_{\exp} \mathbf{v}_1^2}{2} - V_{ext}(\mathbf{r}_1) + \frac{M_{osc} \mathbf{v}_{osc}^2}{2} - \frac{k \cdot \mathbf{r}_{osc}^2}{2} + M_{osc}\left(\alpha_{loss} \frac{\mathbf{v}_1^2}{2} - (\alpha_p)_{\exp} \mathbf{v}_1 \mathbf{v}_{osc}\right)$$

naturally contains *these experimental constants* since we intended just to "make ends meet" with the conservation laws in the previous, rather successful theory (38)-(40). If such a Lagrangian modifies the kinetic terms, it is not a "fault" of the fundamental constant physics or a fault of mathematics. It is a failure of the idea to use a self-action (kinetic) Lagrangian. This idea has never had physical grounds. The physically correct idea is the idea of *permanent interaction* with other bodies. Indeed, in the first particle equation (1), apart from the external force, there is an additional term: it is an *interaction* force rather than a self-action or friction one. As this interaction exists always, we deal in fact with a compound system, so the CIRC approach must be applied for self-consistency. There are no mathematical problems in such an approach and its results agree with experiments. Hence, the self-action idea and its inevitable renormalization ideology should be definitely rejected as misleading physically and wrong mathematically.

Although we know it from the beginning, we can even figure out from the "renormalized" solutions that our particle 1 is coupled to the oscillator *permanently* in our mechanical problem in case of a welded system. Mathematically it is manifested by the oscillating addendum in (48) to the smooth particle-1 *trajectory* (38). It follows from the fact that an oscillating addendum will exists even after the external force is switched off ($t > t_F$) – it will be the term $\propto r_{\max} \cos(\omega_{\exp} t + \varphi_F)$ (as in (19)) where $r_{\max} \neq 0$ and $\varphi_F$ are determined with the matching to the solution (48) at $t = t_F$. In other words, such an oscillating term will be present even in absence of external force and it would be correct if we understood it as a result of coupling to *always present* (somewhere) oscillator (kind of the vacuum fluctuation effect).

It is rather natural then to admit that an always-interacting particle is just a part of a compound system. In other words, a compound system is a synonym of an always-interacting particle. But if it is so, then the observable fundamental constants of the system (masses, frequencies, equation coupling constants) are the quasi particle constants! First, the average inertial constant of our body is the total mass $M_{tot}$. Reasoning so, we arrive at assigning the correct physical meaning to the measured $(M_1)_{\exp}$ and to understanding that our shell is not a rigid body but bound particle 1 with $M_1$. Next, the oscillator in a compound system always represents some *relative* (internal) motion. This understanding signifies that numerically $M_{osc} = \mu$ in our simplest case of one oscillator (one proper frequency system).

As soon as we agree that in fact we deal with a compound system, we can develop the corresponding mechanics of it. Indeed, if acceleration of our body excites an oscillator, it is logical to suppose that this body is simply fixed to one of the oscillator "ends", as in Fig. 1 or Fig. 2. This is a typical compound system whose observable "constants" are the *quasi particle ones*.

### 3.2. Another way of theory formulation: the true self-consistent theory without self-action

Having achieved a new insight in our physical situation, we can easily describe the system CI motion and the system internal energy pumping due to external force action:

1) We rebuild equation (38): we write the CI equation with $\mathbf{R}_{CI}$, the total mass $M_{tot} = (M_1)_{\exp}$ and the external force $\mathbf{F}_{ext}(\mathbf{r}_1)$ (4). The external force argument expressed via CIRC variables $\mathbf{r}_1 = \mathbf{R}_{CI} + \varepsilon \cdot \mathbf{r}_{osc}$ will correctly take into account the fact that the force acts on particle-1 rather than on CI. The time average of $\mathbf{r}_1(t)$ is as smooth as $\mathbf{R}_{CI}(t)$ and it is the only correct understanding of the Newtonian mechanics – it is a mechanics of the CI of compound bodies. (In the quantum approach such an averaging corresponds to the inclusive picture.)

2) We transform the oscillator equation (39) in the following way: in place of (known) particle acceleration $\dot{\mathbf{v}}_1(t)$ we write the (known) external force $\varepsilon \cdot \mathbf{F}_{ext}(\mathbf{r}_1)/\mu$ as a pumping source (5) and understand $\mathbf{r}_{osc}$ as a relative coordinates in a compound system.

Such a formulation is natural and effective both for kit-made and especially for "welded" mechanical systems as it uses *the available experimental (fundamental) constants* without logical and mathematical difficulties. Its Lagrangian is the following:

$$L_{CIRC} = \frac{M_{tot}\mathbf{V}_{CI}^2}{2} - V_{ext}\left(\overbrace{\mathbf{R}_{CI} + \varepsilon \cdot \mathbf{r}_{osc}}^{\mathbf{r}_1}\right) + \frac{\mu \mathbf{v}_{osc}^2}{2} - V(\mathbf{r}_{osc}). \quad (49a)$$

Proceeding from (49a), we obtain good equations (4)-(5) with correct energy conservation. In case of uniform external force, we obtain immediately the exact rather than perturbative solutions.

We can say that in the CIRC approach the quasi-particle equation coupling occurs via a potential term $V_{ext}(\mathbf{r}_1)$ rather than via kinetic ones. When $V_{ext} = 0$, the "mechanical" and the "wave" equations are totally decoupled.

Now we may ask our experimentalists to use a strong pulse ($\omega_{exp} t_F \ll 1$) external force to "measure" $M_1$ and thus $M_2 = (M_1)_{ext} - M_2$. Then we obtain $M_{osc} = \frac{M_1 M_2}{M_1 + M_2} = \mu$, $k = \mu \cdot (\omega_{exp})^2$, and $\varepsilon = \frac{M_2}{M_{tot}}$. From the oscillating part of the measured trajectory $\mathbf{r}_1(t) = \mathbf{R}_{CI}(t) + \varepsilon \cdot \mathbf{r}_{osc}(t) \; \forall t \geq t_F$ we obtain the true oscillator amplitude $r_{max}$. Then we can verify that the energy conservation law holds: the maximum potential energy $k(r_{max})^2/2$ of oscillator is the total internal energy, while $(M_1)_{exp}\langle \mathbf{v}_1 \rangle^2/2$ is the CI energy, and their sum equals the total energy $M_1(\mathbf{v}_1)_{init}^2/2$ initially obtained from the external force during the push.

The Lagrangian (49a) is *the best alternative* to the wrong trial Lagrangian (41) with its inevitable renormalizations. The CIRC formulation should also be recognized as a more fundamental one (the primary) even with respect to the "elementary particle" formulation (1)-(2) because it is not always possible to represent a compound system as some separate bodies coupled with massless springs.

When our welded system is not trivial (think of a ball made entirely of rubber, for example) we will suppose that its internal motion can be represented as a superposition of independent elementary motions (oscillations and possibly rotations). In this case instead of one term $\varepsilon \cdot \mathbf{r}_{osc}$ and one oscillator Lagrangian in (49a) there will be a sum $\sum_k \varepsilon_k \cdot (\mathbf{r}_{osc})_k$ and $\sum_k (L_{osc})_k$, where $k$ numerates different elementary "oscillator" modes:

$$L_{CIRC} = \frac{M_{tot}\mathbf{V}_{CI}^2}{2} - V_{ext}\left(\overbrace{\mathbf{R}_{CI} + \sum_k \varepsilon_k \cdot (\mathbf{r}_{osc})_k}^{\mathbf{r}_1}\right) + \sum_k \left[\frac{\mu_k (\mathbf{v}_{osc})_k^2}{2} - V(\mathbf{r}_{osc})_k\right] \quad (49b)$$

The "first" particle position, which is nothing but the external force application point, is generally rather fluctuating: $\mathbf{r}_1(t) = \mathbf{R}_{CI}(t) + \sum_k \varepsilon_k \cdot (\mathbf{r}_{osc}(t))_k$ (see APPENDIX for classical mechanical examples).

Now our goal is to show that the electron in QED serves as a particle 1 (the external force application point) and the quantized electromagnetic field (as a set of quantum oscillators) describe the internal degrees of freedom of a compound system (electronium). When $V_{ext} = 0$, the *electronium center of inertia* and the *electromagnetic field* are described with *separate* equations ("particle" and "wave" equations) because they belong to the *separated* variables of *one compound system.* When an external field acts on the electron, the quantum oscillators get pumped. This construction is different in physical meaning from the usual understanding – the electromagnetic field is intrinsic to our compound system and it radically simplifies calculations. The true sequences of this approach are only revealed in the quantum mechanical formulation (QM charge smearing, inelastic processes, inclusive cross sections, see Section 3.3.). The classical picture follows from the QED as the inclusive one [7].

### *3.2.1. Hamilton formulation*

To show how and why our approach is useful for building the correct QED, let us now construct the Hamiltonian corresponding to the mixed variables (20) from the Lagrangian (21). The canonical conjugated momenta are:

$$\mathbf{p}_1' = \frac{\partial L'}{\partial \mathbf{v}_1'} = M_1 \mathbf{v}_1' + M_2 (\mathbf{v}_1' - \mathbf{v}_2'), \tag{50}$$

$$\mathbf{p}_2' = \frac{\partial L'}{\partial \mathbf{v}_2'} = -M_2 (\mathbf{v}_1' - \mathbf{v}_2'), \tag{51}$$

The Hamiltonian is expressed as follows:

$$H' = \frac{(\mathbf{p}_1' + \mathbf{p}_2')^2}{2M_1} + V_{ext}(\mathbf{r}_1) + \frac{(\mathbf{p}_2')^2}{2M_2} + V(\mathbf{r}_r).$$

Dropping the primes and denoting $\mathbf{p}_2'$ as $\mathbf{p}_{osc}$ we obtain the exact Hamiltonian:

$$H = \frac{(\mathbf{p}_1 + \mathbf{p}_{osc})^2}{2M_1} + V_{ext}(\mathbf{r}_1) + \frac{\mathbf{p}_{osc}^2}{2M_2} + V(\mathbf{r}_{osc}). \tag{52}$$

Let us introduce a "non-perturbed" (or the "zeroth" order) Hamiltonian:

$$H^{(0)} = \frac{\mathbf{p}_1^2}{2M_1} + V_{ext}(\mathbf{r}_1) + \frac{\mathbf{p}_{osc}^2}{2M_2} + V(\mathbf{r}_{osc}). \tag{53}$$

The "zeroth" order Hamiltonian $H^{(0)}$ describes two "independent" systems (or subsystems). Comparing (52) and (53) we see that the "subsystem" interaction is "switched on" by "enlarging" the "first particle" momentum $\mathbf{p}_1$ by the vector $\mathbf{p}_{osc}$. This "switching on" is fully analogical to the electromagnetic field coupling in QED in the frame of self-action ansatz $\mathbf{p} \to \mathbf{p} - (e/c)\mathbf{A}_{rad}$. The dimension of the dynamical variable $(e/c)\mathbf{A}_{rad}$ is a momentum (a kinetic term!). (Moreover, in CED the Hamilton formulation is not used "to the end" - it is rapidly replaced with the Lagrange formulation with the equations of the second rather than the first order.)

Although $\mathbf{A}_{rad}$ is usually considered as a set of canonical "coordinates", for harmonic oscillators there is a symmetry between canonical coordinates and canonical momenta. So the standard QED formulation is in fact a theory in the mixed variables and *this* is the true reason of arising corrections to the fundamental (or phenomenological) constants. The fundamental constants in (53) are *observable*, just like those from (38), (39). They are not "bare" in the sense of being "non observable", as it is said to justify QFT renormalization practice. They appear first in non-perturbed QED Hamiltonian which is similar to $H^{(0)}$ (53) or in the non-perturbed QED equations, which are similar to (38) and (39) for the welded system. In fact, there are no "bare" constants; there is simply a wrong self-action ansatz (in the Hamilton formulation it is $\mathbf{p} \to \mathbf{p} - (e/c)\mathbf{A}_{rad}$) that does not describe the experimental data without sacrificing physical sense and mathematical strictness in course of calculations. With the good – "interaction" ansatz, no corrections to the previously measured (fundamental) constants arise. The corresponding correct Hamiltonian is obtained elementarily from the Lagrangian (49a) by expressing the kinetic energy terms with in terms of momenta ($p^2/2m$) accompanied with changing the signs at the potential energy terms.

Comparison of (52) and (53) means that actually the electron and the quantized EM field, being coupled permanently, form a compound system. I called it an *electronium* for certainty [7]. Our goal is to construct its CIRC Lagrangian or Hamiltonian. Actually, only quantum mechanical treatment provides the correct results (the quantum mechanical charge smearing, for example) and the correct classical limit as the inclusive picture.

### 3.3. Towards correct QED

In QED the transversal vector potential $\mathbf{A}_{rad}$ is considered as canonic coordinates rather than momenta. Can $(e/c)\mathbf{A}_{rad}$ be presented as a sum of oscillator momenta? Yes, it can. Let us represent it as a sum over the wave vectors $\mathbf{k}$ (which just numerate the independent modes) and introduce the momenta $\mathbf{p_k}$ in an evident way:

$$-\frac{e}{c}\mathbf{A}_{rad} = -\sum_{\mathbf{k}} \frac{e}{c}\tilde{\mathbf{A}}_{\mathbf{k}} = \sum_{\mathbf{k}} \mathbf{p_k}, \quad \mathbf{p_k} = -\frac{e}{c}\tilde{\mathbf{A}}_{\mathbf{k}}.$$

The Hamilton equations for an oscillator are: $\dot{\mathbf{x}} = \mathbf{p}/m_{osc}$; $\dot{\mathbf{p}} = -\tilde{\alpha}\mathbf{x}$ where $\tilde{\alpha}$ denotes the oscillator elasticity and $\mathbf{x}$ the oscillator canonic coordinate (= relative coordinate in electronium). What are the oscillator coordinates conjugated to the momenta $\mathbf{p_k}$? From the vector potential definition it follows: $(e/c)\dot{\tilde{\mathbf{A}}}_{\mathbf{k}}(t) = -e\mathbf{E}_{\mathbf{k}}(t)$. Then

$$\dot{\mathbf{p}}_{\mathbf{k}} = -\frac{e}{c}\dot{\tilde{\mathbf{A}}}_{\mathbf{k}} = -(-e\mathbf{E}_{\mathbf{k}}(t)) = -\tilde{\alpha}_{\mathbf{k}}\mathbf{x}_{\mathbf{k}}, \quad \boxed{\mathbf{x}_{\mathbf{k}} = -\frac{e\mathbf{E}_{\mathbf{k}}}{\tilde{\alpha}_{\mathbf{k}}}}, \quad [\tilde{\alpha}_{\mathbf{k}}] = \frac{Force}{Length}.$$

Thus each relative electronium coordinate $\mathbf{x}_{\mathbf{k}}$ is proportional to its oscillator field tension $\mathbf{E}_{\mathbf{k}}(t)$. We should pick up $\tilde{\alpha}_{\mathbf{k}}$ in the way to respect the correct units of $\mathbf{x}_{\mathbf{k}}$ (meters).

$$\underline{\underline{\dot{\mathbf{p}}_{\mathbf{k}} = -\tilde{\alpha}_{\mathbf{k}}\mathbf{x}_{\mathbf{k}}}}, \quad \underline{\underline{\dot{\mathbf{x}}_{\mathbf{k}} = \frac{c^2\mathbf{k}^2}{\tilde{\alpha}_{\mathbf{k}}}\mathbf{p_k}}}, \quad \tilde{m}_{osc.}(\mathbf{k}) = \frac{\tilde{\alpha}_{\mathbf{k}}}{c^2\mathbf{k}^2} > 0, \quad \tilde{\omega}_{osc.}(\mathbf{k}) = \sqrt{\frac{\tilde{\alpha}_{\mathbf{k}}}{\tilde{m}_{osc.}(\mathbf{k})}} = c|\mathbf{k}|.$$

The oscillator equations coincide with the well-known ones:

$$\ddot{\mathbf{x}}_{\mathbf{k}} + \tilde{\omega}_{\mathbf{k}}^2 \mathbf{x}_{\mathbf{k}} = 0 \quad or \quad \ddot{\mathbf{E}}_{\mathbf{k},\lambda}(t) = -c^2\mathbf{k}^2\mathbf{E}_{\mathbf{k},\lambda}.$$

The non relativistic QED without self-action, divergences, and renormalizations, based on these findings (that I call the "interaction ansatz"), can be built with choosing $\tilde{\alpha}_{\mathbf{k}} = m_e c^2 \mathbf{k}^2$ and limiting the photon wave vectors to $k_{max} \approx m_e c / \hbar$. For example, the non-relativistic Hamiltonian of interaction of the electronium with particle 3 (the latter acts on the electron via potential $V(\mathbf{r}_3 - \mathbf{r}_e)$) is given with the formula:

$$H_{tot} = \frac{\mathbf{p}_3^2}{2M_3} + \frac{\mathbf{P}_{CIe}^2}{2m_e} + V(\mathbf{r} + \sum_{\mathbf{k},\lambda}^{k_{max}} \frac{-e \cdot \mathbf{E}_{\mathbf{k},\lambda}}{m_e c^2 \mathbf{k}^2}) + H_{osc}. \qquad (54)$$

(Here we suppose that particle 3 is not coupled to the quantized EMF directly, unlike the electron.) Compare it with the Lagrangian (49b). Here $H_{osc}$ is the Hamiltonian of the quantized electromagnetic field represented as a sum of quantum oscillators. The Hamiltonian (54) describes scattering with inevitable bremsstrahlung as well as the Lamb shift. It replaces the non-relativistic Hamiltonian with the self-action ansatz. The electronium negative charge is smeared quantum mechanically and is described with the corresponding form-factors [7]. In the first Born approximation calculated in the center of mass of the projectile and electronium the cross sections are given with the formulas:

$$\frac{d\sigma_n^{n'\,p'}(\theta)}{d\Omega} = \frac{m^2}{4\pi^2\hbar^4}\frac{p'}{p}\left|\int V(\mathbf{r})e^{-i\mathbf{qr}}d^3r\right|^2 \cdot \left|f_n^{n'}(\mathbf{q})\right|^2 \qquad (55)$$

$$f_n^{n'}(\mathbf{q}) = \int \exp\left(-i\mathbf{q}\sum_{\mathbf{k},\lambda} e\frac{\mathbf{e}_{\mathbf{k},\lambda}Q_{\mathbf{k},\lambda}}{m_e c^2 \mathbf{k}^2}\right)\prod_{\mathbf{k},\lambda}\chi_{\mathbf{k},\lambda}\chi_{\mathbf{k},\lambda}^* dQ_{\mathbf{k},\lambda}, \qquad (56)$$

with $m = M_3 m_e/(M_3 + m_e)$; $\mathbf{p} = m\mathbf{v}$; $\hbar\mathbf{q} = \mathbf{p}' - \mathbf{p}$; $p' = \sqrt{p^2 - 2m(E_{n'} - E_n)}$. $\qquad (57)$

The elastic form-factor of a free electronium $f_0^0(\mathbf{q})$ is equal to zero, as it should be:

$$f_0^0(\mathbf{q}) \propto \exp\left(-\int_{k_{min}}^{k_{max}}(\mathbf{qe})^2\frac{4\pi e^2}{m_e^2 k^3 c^3 \hbar}\frac{k^2 dk do}{(2\pi)^3}\right) = \exp\left(-\varsigma(q)\ln\frac{\omega_{max}}{\omega_{min}}\right)\bigg|_{\omega_{min}\to 0} \to 0. \qquad (58)$$

That means the negative charge in a free electronium is smeared quantum mechanically over the whole space due to vacuum field fluctuations, so it is never pointlike (the picture is only valid for elastic process). Any kind of binding electronium (in atom, in magnetic field or behind a diaphragm) makes the smearing size finite ($\omega_{min} > 0$ or $k_{min} > 0$) [7]. These physical effects are known from T. Welton's publication [8] but unfortunately they are not considered seriously (they are presented as a qualitative approach).

In the momentum space the elastic form-factor, if non zero, serves as a *natural regularization factor* as it tends rapidly to zero when $|\mathbf{q}| \to \infty$. It is useful in higher orders of relativistic calculations in external fields, which are "switched on" as usually: $P \to P + (e/c)A_{ext}(x_e)$ and provide the gauge invariance.

It is easy to verify that all inelastic electronium form-factors $f_n^{n'}$ with finite number of final photons are also equal to zero, as it should be. The totally inclusive cross section is different from zero and it is reduced accurately enough to the "mechanical" cross section due to extremely weak dependence of $p'$ and $q$ on $E_{n'} - E_n$ and due to the sum rule $\sum_{n'}\left|f_n^{n'}\right|^2 = \left|f\,f^+\right|_n^n = 1$, just as in the atomic case [7]:

$$d\sigma_{inclusive}(\mathbf{q}) \approx d\sigma_{mechanical} = \frac{m^2}{4\pi^2\hbar^4}\left|\int V(\mathbf{r})e^{-i\mathbf{qr}}d^3r\right|^2 d\Omega. \qquad (59)$$

We see that the inclusive picture "corresponds" to scattering from the electronium CI as if the target were "pointlike", without internal degrees of freedom, and situated at the CI$_e$. The energy spent on excitations $E_{n'} - E_n$ in our case is much smaller than that spent on the whole target acceleration. The pioneering experimentalists have dealt with the *inclusive* cross sections rather than with the elastic ones because they could not distinguish different inelastic processes. This fact explains why the notions of point-like elementary electrons and nuclei have appeared and are still so widespread. Having created the notion of a neutral pointlike particle and then by analogy a pointlike charge, the researchers have separated the electromagnetic field from charges - they write the Maxwell equations without sources, they study the electromagnetic waves at far distances from charges, etc. In our quantum electronium, on the contrary, the charge is always smeared and the photons are just excited states of electronium. In other words, instead of saying that the soft radiation has the classical nature (where the charge is pointlike and classical, its recoil neglected), it is correct to say that the classical (soft) radiation is the inclusive quantum mechanical result (the charge is quantum mechanical, recoil taken into account).

All classical results ("mechanical" cross sections and even classical trajectories as series of successive scatterings) are obtained now due to *taking into account* the radiation processes (59), as in factually inclusive experiments, rather than due to neglecting them (i.e., the term $(e/c)\mathbf{A}_{rad}$ in the self-action QED).

Perturbative treatment of the term $\sum e\mathbf{E}_{\mathbf{k},\lambda}/m_e c^2 \mathbf{k}^2$ in (54) in scattering problems leads to the infrared divergence. This is seen from the development of the electronium elastic form-factor (58) in powers of $e^2$. Hence the quantum mechanical smearing effect *cannot* be obtained by the perturbation theory when the initial and final particles are free. It should be taken into account in the first turn.

The partially inclusive cross section – with summation on the soft photon energies from zero to $E_l$ – is also different from zero and it is reduced accurately enough to the "mechanical" cross section multiplied by the partially inclusive electronium form-factor depending on $E_l$ (kind of Sudakov's form-factor):

$$\sum_{n'}^{E_l} \left|f_n^{n'}\right|^2 = F(E_l) \leq 1.$$

If particle 3 is charged too, it is natural to assume that it is coupled to *its own* quantized EMF, and the cross sections are expressed via a product of the two form-factors, one per charge [7].

It is evident that the initial electronium state cannot be the "ground state" – a state with no initially excited photons (it is extremely difficult to prepare such a state). In fact, it is nearly always a mixed (or in a superposition) state, and we have in (59), to be exact, to average over all possible initial electronium states too.

The Hamiltonian (54) provides also the Lamb shift estimation without problem (shown for the first time in [8]).

In case of cavity QED, the photon spectrum will have a threshold.

The trial relativistic Hamiltonian of the Novel QED in the Heisenberg picture can be given with the expression [7]:

$$H = \int d^3 P \sum_{\substack{c=electron,\\positron}} \left\{\pi_c(\mathbf{P},t)\gamma^0 (i\gamma\mathbf{P}\eta_c + m_e)u_c(\mathbf{P},t) + H_{osc\mathbf{P},c}\right\} + \frac{1}{2}\int d^3 R_1 \int d^3 R_2 \frac{j^0(\mathbf{R}_1,t)j^0(\mathbf{R}_2,t)}{4\pi|\mathbf{r}_1-\mathbf{r}_2|} \quad (60)$$

It resembles the Coulomb-gauge one (see formula (A.25) in [9]), but here $\mathbf{P}$ is the electronium (or better "fermionium") CI momentum. Each free CI Hamiltonian is accompanied with its own photon Hamiltonian $H_{osc\mathbf{P},c}$ in order to represent together a free compound system. The self-action term $\mathbf{j}\cdot\mathbf{A}_{rad}$ is omitted as originating from the wrong self-action ansatz. Instead, the quantum oscillator tensions are "inserted" into the Cartesian coordinates $\mathbf{r}_c = \mathbf{R}_{CIc} - \sum_{\mathbf{k},\lambda} e_c \mathbf{E}_{\mathbf{k},\lambda}/(m_e c^2 \mathbf{k}^2)$ for each charge (electron and positron). The Coulomb interaction responsible both for scattering and bound states of fermioniums is expressed via $CI_c$ and the relative variables of each charge involved:

$$\mathbf{r}_1 - \mathbf{r}_2 = (\mathbf{R}_{CI})_1 - (\mathbf{R}_{CI})_2 - \sum_{\mathbf{k},\lambda}\left[(e\mathbf{E}_{\mathbf{k},\lambda})_1 - (e\mathbf{E}_{\mathbf{k},\lambda})_2\right]/(m_e c^2 \mathbf{k}^2). \quad (61)$$

This provides each fermionium with its own form-factor due to its own oscillator field influence. Factually the expression $\mathbf{r}_c = \mathbf{R}_{CIc} - \sum_{\mathbf{k},\lambda} e_c \mathbf{E}_{\mathbf{k},\lambda} / (m_e c^2 \mathbf{k}^2)$ in our theory is of a very significant meaning. It replaces the wrong "minimal coupling" $D_\mu = \partial_\mu + eA_\mu$ (self-action) ansatz with its inevitable renormalizations.

No infrared divergences arise here since any scattering becomes formally a *potential scattering of compound quantum mechanical systems* (fermioniums) with inevitable exciting their "internal" degrees of freedom (photons). The obligatory inclusive consideration in such a theory yields the results corresponding to the inclusive experiments.

No ultraviolet divergences (UV) arise either since, first, no fermion-radiation self-action is introduced in our fermionium model, and second, the self-energy fermion loops originating from the four-fermion Coulomb interaction in higher orders vanish in scattering problems due to vanishing the elastic form-factors of the free fermionium at each vertex. In bound states the fermionium form-factor is different from zero but it makes the loop contributions finite and rather small thanks to its significant regularization property. Thus, the problems of IR and UV divergences do not exist in the Novel QED thanks to using the notion of electronium. No bare constants are introduced, no renormalization is necessary, and there is no such a feature as the Landau pole.

The correct value of anomalous magnetic moment is obtained only from the relativistic formulation because it is a spin-dependent relativistic quantum effect. T. Welton's estimations demonstrated arising and correct order of the anomalous magnetic moment but the opposite sing since these estimations were non-relativistic.

The external fields (including single photon case) are not included in (60) for simplicity but they can be added in a usual for the Coulomb gauge way (see APPENDIX, for example).

An interesting alternative to study is to consider, along with photons, the electron-positron pairs as possible excitations of the "internal" degrees of freedom of electronium. The electron-positron pair Hamiltonians can be added to the electronium CI Hamiltonian as separated "internal" variable Hamiltonians. The distance (61) should also be modified in order to take the additional degrees of freedom into account.

## 4. CONCLUSION

As we could see, the self-consistent theory can be formulated in the frame of the "interaction" ansatz where fermions and bosons represent the CI and relative motions of compound systems.

I believe that the other "gauge" field theories should be reformulated in the same way: the corresponding self-action terms (gauge covariant derivative $D_\mu = \partial_\mu + eA_\mu$) should be replaced with the fermionium CI free motion derivative, the "gauge" field tensions should be "inserted" in the fermion Cartesian coordinates to describe the relative (internal) degrees of freedom and symmetries of the corresponding compound "fermioniums". Then, for example, free quarks and gluons will not exist in the theory in full agreement with non-existence of free electron and photon in our electronium dynamics. The particle theory will recover a phenomenological character instead of being highly limited with mathematical constrains (gauge invariance, renormalizability, etc.). I invite researchers to explore this direction of the "elementary" particle theory where the "elementary" are excitation modes of compound systems.

**APPENDIX**

**Comparison of non-relativistic CED and Novel CED in the Lagrangian formulation**

The one-particle NCED Lagrangian for a charge $q$ can be written in the following way:

$$L = L_{free} + L_{int},$$

$$L_{free} = \frac{1}{2}m\dot{\vec{R}}^2 + \frac{1}{2}\int d^3k \left[ \dot{\vec{a}}^*_{rad}(\mathbf{k})\dot{\vec{a}}_{rad}(\mathbf{k}) - \ddot{\vec{a}}^*_{rad}(\mathbf{k})\ddot{\vec{a}}_{rad}(\mathbf{k})/c^2 k^2 \right], \tag{A1}$$

$$L_{int} = q\left[ \varphi_{ext}(\vec{r}) + \frac{\dot{\vec{r}}}{c} \cdot \vec{A}_{ext}(\vec{r}) \right], \tag{A2}$$

$$\vec{r} = \vec{R} + q\int d^3k\, \varepsilon(k) \left[ \dot{\vec{a}}_{rad}(\mathbf{k}) - \dot{\vec{a}}^*_{rad}(\mathbf{k}) \right]. \tag{A3}$$

I use small letters $\vec{a}_{rad}(\mathbf{k})$ for the radiated field Fourier amplitudes in order not to mix them with the radiated field in the coordinate space $\vec{E}_{rad}(\mathbf{x},t)$. Their time derivatives $\dot{\vec{a}}_{rad}(\mathbf{k})$ determine the corresponding electric field Fourier amplitudes. (The radiated field $\vec{E}_{rad}(\mathbf{x},t)$, as well as the total field created by a given charge, acts on another charge, if any, and depends not only on $\dot{\vec{a}}_{rad}(\mathbf{k})$ but also on the distance from one charge to another.)

Now let us derive the "force" part of the Lagrangian equations. I denote $\vec{Q}$ any generalized coordinate involved, i.e., the center of inertia and field variables. It is convenient also to use $\vec{r}$ for physical (force application point) and mathematical reasons (shorter expressions).

$$\frac{\partial L_{int}}{\partial Q_i} = \left( -q\frac{\partial \varphi_{ext}}{\partial r_j} + q\dot{\vec{r}}\frac{\partial \vec{A}_{ext}}{\partial r_j} \right)\frac{\partial r_j}{\partial Q_i},$$

$$\frac{\partial r_j}{\partial Q_i} = f(Q,\vec{k})\cdot\delta_{ij}; \quad f(Q,\vec{k}) = \begin{cases} 1, & \text{if } \vec{Q}=\vec{R} \\ \pm q\cdot\varepsilon(k), & \text{if } \vec{Q}=\vec{a} \text{ or } \vec{a}^* \end{cases}$$

Hence we obtain:

$$\frac{\partial L_{int}}{\partial Q_i} = \left( -q\frac{\partial \varphi_{ext}}{\partial r_i} + q\dot{\vec{r}}\frac{\partial \vec{A}_{ext}}{\partial r_i} \right) f(Q,\vec{k}). \tag{A4}$$

Next:

$$\frac{\partial L_{int}}{\partial \dot{Q}_i} = q\vec{A}_{ext}\frac{\partial \dot{\vec{r}}}{\partial \dot{Q}_i}; \quad \frac{d}{dt}\frac{\partial L_{int}}{\partial \dot{Q}_i} = q\dot{\vec{A}}_{ext}\frac{\partial \dot{\vec{r}}}{\partial \dot{Q}_i} + q\vec{A}_{ext}\frac{d}{dt}\frac{\partial \dot{\vec{r}}}{\partial \dot{Q}_i},$$

$$\frac{\partial \dot{r}_j}{\partial \dot{Q}_i} = f(Q,\vec{k}) \cdot \delta_{ij}; \quad f(Q,\vec{k}) = \begin{cases} 1, & \text{if } \dot{Q} = \dot{\vec{R}} \\ \pm q \cdot \varepsilon(k), & \text{if } \dot{Q} = \ddot{\vec{a}}, \ddot{\vec{a}}^* \end{cases} \quad ; \quad \frac{d}{dt}\frac{\partial \dot{\vec{r}}}{\partial \dot{Q}_i} = 0.$$

Thus we obtain:

$$\frac{d}{dt}\frac{\partial L_{int}}{\partial \dot{Q}_i} = q\dot{\vec{A}}_{ext}\frac{\partial \dot{\vec{r}}}{\partial \dot{Q}_i} = q\left(\dot{\vec{A}}_{ext}\right)_i f(Q,\vec{k}). \tag{A5}$$

Put together, they read:

$$\frac{d}{dt}\frac{\partial L_{int}}{\partial \dot{Q}_i} - \frac{\partial L_{int}}{\partial Q_i} = q\left(\dot{A}_i^{ext} + \frac{\partial \varphi_{ext}}{\partial r_i} - \dot{\vec{r}}\frac{\partial \vec{A}_{ext}}{\partial r_i}\right) f(Q,\vec{k}). \tag{A6}$$

The full-time derivative:

$$\dot{A}_i^{ext} = \frac{\partial A_i^{ext}}{\partial t} + \left(\dot{\vec{r}} \cdot \nabla_{\vec{r}}\right) A_i^{ext}.$$

Hence:

$$\frac{d}{dt}\frac{\partial L_{int}}{\partial \dot{Q}_i} - \frac{\partial L_{int}}{\partial Q_i} = q\left(\frac{\partial A_i^{ext}}{\partial t} + \frac{\partial \varphi_{ext}}{\partial r_i} + \left(\dot{\vec{r}} \cdot \nabla_{\vec{r}}\right) A_i^{ext} - \dot{\vec{r}}\frac{\partial \vec{A}_{ext}}{\partial r_i}\right) f(Q,\vec{k}),$$

or in the vector notations:

$$\frac{d}{dt}\frac{\partial L_{int}}{\partial \dot{\vec{Q}}} - \frac{\partial L_{int}}{\partial \vec{Q}} = q \cdot f(Q,\vec{k})\left[\frac{\partial \vec{A}_{ext}}{\partial t} + \nabla_{\vec{r}}\varphi_{ext} + \left(\dot{\vec{r}} \cdot \nabla_{\vec{r}}\right)\vec{A}_{ext} - \nabla_{\vec{r}}\left(\dot{\vec{r}} \cdot \vec{A}_{ext}\right)\right]. \tag{A7}$$

Using the vector dot product property:

$$\nabla(\vec{A} \cdot \vec{B}) = (\vec{A} \cdot \nabla)\vec{B} + (\vec{B} \cdot \nabla)\vec{A} + \vec{A} \times (\nabla \times \vec{B}) + \vec{B} \times (\nabla \times \vec{A}),$$

where $\vec{B} = \dot{\vec{r}}$ and $\vec{A} = \vec{A}_{ext}(\vec{r},t)$ we obtain:

$$\frac{d}{dt}\frac{\partial L_{int}}{\partial \dot{\vec{Q}}} - \frac{\partial L_{int}}{\partial \vec{Q}} = q \cdot f(Q,\vec{k})\left[\frac{\partial \vec{A}_{ext}}{\partial t} + \nabla_{\vec{r}}\varphi_{ext} - \dot{\vec{r}} \times \left(\nabla_{\vec{r}} \times \vec{A}_{ext}\right)\right],$$

or:

$$\frac{d}{dt}\frac{\partial L_{int}}{\partial \dot{\vec{Q}}} - \frac{\partial L_{int}}{\partial \vec{Q}} = -q \cdot f(Q,k)\left[\vec{E}_{ext}(\vec{r}) + \dot{\vec{r}} \times \vec{B}_{ext}(\vec{r})\right], \tag{A8}$$

$$\vec{Q} = \vec{R}, \vec{a}, \vec{a}^*; \quad f(Q,k) = \begin{cases} 1, & \text{if } \vec{Q} = \vec{R}, \\ \pm q \cdot \varepsilon(k), & \text{if } \vec{Q} = \vec{a}, \vec{a}^*. \end{cases}$$

**Both** "particle" ($\vec{R}$) and "field" or oscillator ($\vec{a}$) equations have the same external (Lorentz) force on the right-hand side (in the square brackets). Thus the total external force work splits naturally into two parts:

the CI and "internal" energy modifications. (In case of an external magnetic field, it only helps convert the particle kinetic energy into radiation.)

Now let us look at the CED equations in the Lorentz gauge:

$$\ddot{\vec{A}}^*(\mathbf{k}) + \omega_k^2 \vec{A}^*(\mathbf{k}) \propto \vec{j}^*(\mathbf{k}) \tag{A9}$$

In CED the current is determined not only with external fields but also with the "proper" charge field. Although invisible from the Lagrangian formulation (where one thinks the beautiful Noether theorem excludes any non-physical possibilities), the self-action leads to senseless mathematically and physically expressions – infinities – in case of point-like particles. In plasma and other macroscopic systems with smooth (averaged) charge densities and currents this problem does not exists. But in model of point-like electron the exact CED calculations are impossible. That is why H. Lorentz had to discard the infinite addendum to the mechanical equation. It is erroneously called the mass renormalization but it is merely discarding and postulating a new equation, as I showed it in the main text. The remaining finite term - jerk $\dddot{\vec{r}}$ - still leads to non-physical, runaway solutions and hence cannot be treated exactly. At most it is treated perturbatively but usually it is discarded too.

Thus the microscopic CED is factually reduced to and practiced in two limiting situations:

1) known current, unknown filed: $L_{int} = j_{ext} \cdot A$, or
2) known field and unknown current (mechanics): $L_{int} = j \cdot A_{ext}$.

Let us however obtain the point-like charge current Fourier image:

$$\vec{j}(\vec{r}) = \int d^3 k\, \vec{J}(\mathbf{k}) e^{i\mathbf{kr}}; \quad \vec{J}(\mathbf{k}) = \frac{1}{(2\pi)^3} \int d^3 r\, \vec{j}(\mathbf{r}) e^{-i\mathbf{kr}},$$

$$\vec{j}(\mathbf{r}) = q\dot{\vec{r}} \delta^3(\mathbf{r} - \vec{r}(t)), \quad \vec{J}(\mathbf{k},t) = \frac{1}{(2\pi)^3} \int d^3 r\, \vec{j}(\mathbf{r}) e^{-i\mathbf{kr}} = \frac{q\dot{\vec{r}}(t)}{(2\pi)^3} e^{-i\mathbf{kr}(t)}. \tag{A10}$$

So the field equation in CED for a point-like charge reads:

$$\ddot{\vec{A}}(\mathbf{k}) + \omega_k^2 \vec{A}(\mathbf{k}) \propto q\dot{\vec{r}}(t) e^{-i\mathbf{kr}(t)}. \tag{A11}$$

It is different from my filed equations due to presence of an exponential factor and including in $\vec{A}(\mathbf{k})$ the magnetic field of a uniformly moving charge. Let us use a dipole approximation in CED. Then the exponential equals to unity: $e^{-i\mathbf{kr}(t)} \approx 1$. The dipole approximation is anyway valid in the non-relativistic case $v/c \ll 1$. The field equations become:

$$\ddot{\vec{A}}(\mathbf{k}) + \omega_k^2 \vec{A}(\mathbf{k}) \propto q\dot{\vec{r}}(t). \tag{A12}$$

For the radiated part $\vec{a}_{rad}(\mathbf{k})$ of the total vector-potential harmonic $\vec{A}(\mathbf{k})$, which is also determined in CED with acceleration rather than velocity, we obtain:

$$\ddot{\vec{a}}_{rad}(\mathbf{k}) + \omega_k^2 \vec{a}_{rad}(\mathbf{k}) \propto q\ddot{\vec{r}}(t). \tag{A13}$$

Now I borrow the expression of the charge acceleration from the Newton equation in case the current is determined with external fields solely (i.e., no self-action, no jerk contribution in (**I**2), situation 2). I denote the 31corresponding solution of mechanical equations as $\vec{r}^{(0)}(t)$. The acceleration is proportional to the Lorentz force:

$$\dddot{\vec{a}}_{rad}(\mathbf{k}) + \omega_k^2 \dot{\vec{a}}_{rad}(\mathbf{k}) \propto q^2 \left[ \vec{E}_{ext}(\vec{r}^{(0)}(t)) + \dot{\vec{r}}^{(0)}(t) \times \vec{B}_{ext}(\vec{r}^{(0)}(t)) \right]. \tag{A14}$$

As we can see, the "simplified" CED wave equations (A14) are similar to mine (A8) obtained without simplifications. Therefore the NCED results on radiation coincide in the main with the CED ones. (As the polarization vector directions are determined with the conditions $\vec{k} \| \vec{r}$, $\vec{a}_{rad} \perp \vec{r}$, instead of (A14) one must use its projections on the unit polarization vectors.)

Actually the formula (A14) may serve as a hint that an electron is a part of the field oscillators in CED.

Now let me cite a passage from the Feynman's lectures on Physics:

### 28-5 Attempts to modify the Maxwell theory

We would like now to discuss how it might be possible to modify Maxwell's theory of electrodynamics so that the idea of an electron as a simple point charge could be maintained. Many attempts have been made, and some of the theories were even able to arrange things so that all the electron mass was electromagnetic. But all of these theories have died. It is still interesting to discuss some of the possibilities that have been suggested—to see the struggles of the human mind.

We started out our theory of electricity by talking about the interaction of one charge with another. Then we made up a theory of these interacting charges and ended up with a field theory. We believe it so much that we allow it to tell us about the force of one part of an electron on another. Perhaps the entire difficulty is that electrons do not act on themselves; perhaps we are making too great an extrapolation from the interaction of separate electrons to the idea that an electron interacts with itself. Therefore some theories have been proposed in which the possibility that an electron acts on itself is ruled out. Then there is no longer the infinity due to the self-action. Also, there is no longer any electromagnetic mass associated with the particle; all the mass is back to being mechanical, but there are new difficulties in the theory.

We must say immediately that such theories require a modification of the idea of the electromagnetic field. You remember we said at the start that the force on a particle at any point was determined by just two quantities—*E* and *B*. If we abandon the "self-force" this can no longer be true, because if there is an electron in a certain place, the force isn't given by the total *E* and *B*, but by only those parts due to *other* charges. So we have to keep track always of how much of *E* and *B* is due to the charge on which you are calculating the force and how much is due to the other charges. This makes the theory much more elaborate, but it gets rid of the difficulty of the infinity.

So we can, *if we want to*, say that there is no such thing as the electron acting upon itself, and throw away the whole set of forces in Eq. (28.9). However, we have then thrown away the baby with the bath! Because the second term in Eq. (28.9), the term in $\ddot{x}$, is needed. That force does something very definite. If you throw it away, you're in trouble again. When we accelerate a charge, it radiates electromagnetic waves, so it loses energy. Therefore, to accelerate a charge, we must require more force than is required to accelerate a neutral object of the same mass; otherwise energy wouldn't be conserved. The rate at which we do work on an accelerating charge must be equal to the rate of loss of energy per second by radiation. We have talked about this effect before—it is called the radiation resistance. We still have to answer the question: Where does the extra force, against which we must do this work, come from?

arXiv:0811.4416

As we can see from this text, attempts have been made to get rid of self-action but then no additional "force" was introduced into the "mechanical" equations to keep track of the "radiation resistance". In our model, on the other hand, the "resistance" forces are permanently present in equations for $\mathbf{r}(t)$ (by its definition) and are the "returning" forces of oscillators. Thus we do not need a harmful term $\dddot{\mathbf{r}}(t)$ that spoils physics. (Any derivation of the "resistance effects" of this term (integrations by part, etc.) is based on supposition that the mechanical equation have physically reasonable solutions which is not the case for $\dddot{\mathbf{r}}(t)$.)

In case of many charges in NCED their dynamics is determined not only with external fields but also with their interactions. As Feynman said, this makes the theory much more elaborate but it is still possible. I will present the corresponding formulas (equations and Lagrangians, if any) in a separate publication.

For a one-particle dynamics my approach joins self-consistently the both limiting cases 1) and 2) in the frame of a new concept – electronium. The coupling is carried out via "potential" (A2), (A3) rather than kinetic terms. Although harmonics $\vec{a}_\alpha(\mathbf{k},t)$ may be coupled and $\vec{R}_\alpha(t)$ may oscillate, it is not a self-action and it does not create any pathology in solutions. The only non-physical situation might be the electron-proton collapse, which is resolved with QM.

In practice it is a time average $\langle \vec{r}_\alpha(t) \rangle_T$ that has the physical, observable sense in the classical (not QM) approach. In many cases $\langle \vec{r}_\alpha(t) \rangle_T$ is rather close to $\langle \vec{R}_\alpha(t) \rangle_T$. One can derive the dynamics equations even for $\langle \vec{R}_\alpha(t) \rangle_T$. In simplest cases the oscillating part of $\langle \vec{R}_\alpha(t) \rangle_T$ is negligible and the $\langle \vec{R}_\alpha(t) \rangle_T$ dynamics is determined with the external forces solely (a uniform or a nearly uniform external electric or gravitational field, for example). Anyway, we should not give too much sense to this classical picture since instead of time averaging we should use the quantum mechanical inclusive picture. The QM description is more fundamental and the mechanical analogy above is given just for qualitative explanations. In particular, we cannot consider the oscillating modes $\vec{a}_\alpha(\mathbf{k},t)$ to be some sort of "hidden" variables however strong our temptation might be. The pointlike particle is, strictly speaking, the inclusive QM picture. In NCED it corresponds to the averaged (highly fluctuating otherwise) electron coordinate. Highly fluctuating electron coordinate is unusual thing for CED. Dynamics of $\langle \vec{R}_\alpha(t) \rangle_T$ is more "certain" and observable. In particular, in case of small coordinate dispersions with respect to large inter-charge average distances $\langle |\vec{r}_\alpha - \vec{r}_\beta| \rangle_T \approx \left| \langle \vec{R}_\alpha(t) \rangle_T - \langle \vec{R}_\beta(t) \rangle_T \right|$ we obtain ("recover") the usual pointlike charge dynamics. Again, strictly speaking, the electron coordinate "dispersion" around some average $\langle \vec{R}_\alpha(t) \rangle_T$ is determined with QM mechanism although it has a classical mechanical analogue here.